\documentclass[pdflatex,sn-mathphys-num]{sn-jnl}% Math and Physical Sciences Numbered Reference Style
\usepackage{graphicx}%
\usepackage{multirow}%
\usepackage{amsmath,amssymb,amsfonts}%
\usepackage{amsthm}%
\usepackage{mathrsfs}%
\usepackage[title]{appendix}%
\usepackage{xcolor}%
\usepackage{textcomp}%
\usepackage{manyfoot}%
\usepackage{booktabs}%
\usepackage{algorithm}%
\usepackage{algorithmicx}%
\usepackage{algpseudocode}%
\usepackage{listings}%
\usepackage{natbib}
\usepackage{capt-of}
\usepackage{graphicx}

\usepackage{url}
\usepackage{graphicx}% Include figure files
\usepackage{dcolumn}% Align table columns on decimal point
\usepackage{bm}% bold math
\usepackage{multirow}
\usepackage{tabularx}
\usepackage{adjustbox}
\usepackage{hyperref}
\usepackage{natbib}
\usepackage{amssymb}
\usepackage{amsmath}
\usepackage{stackengine}
\usepackage{pgfplots}
\usepackage{adjustbox}
\usepackage{makecell, multirow, tabularx}
\usepackage{array}
\usepackage{booktabs}
\usepackage{mathtools}

\usepackage{float}
\usepackage{stfloats}
\usepackage{enumitem}
\usepackage{caption}
\usepackage{subcaption}
\pgfplotsset{compat=1.18}
\usepackage[utf8]{inputenc}
\usepackage[T1]{fontenc}
\usepackage{mathptmx}
\usepackage{etoolbox}
\usepackage{siunitx} 
\usepackage{adjustbox}\usepackage{multirow}   % multirow cells
\usepackage{xcolor} 
\theoremstyle{thmstyleone}%
\theoremstyle{thmstyletwo}%

\theoremstyle{thmstylethree}%

\begin{document}

%\title[]{Human and Algorithmic Behavior During the 2024 U.S. Presidential Election: Evidence from ERC-20 Stablecoin Transactions}

\title{Early-Warning Signals of Political Risk in Stablecoin Markets: Human and Algorithmic Behavior Around the 2024 U.S. Election}

\author[1]{\fnm{Kundan} \sur{Mukhia}}
%\email{kundanmukhia07@gmail.com}
\author[1]{\fnm{Buddha Nath} \sur{Sharma}}
%\email{bnsharma09@yahoo.com}
\author[1]{\fnm{Salam Rabindrajit} \sur{Luwang}}
%\email{salamrabindrajit@gmail.com}
\author[1]{\fnm{Md.} \sur{Nurujjaman}}
%\email{md.nurujjaman@nitsikkim.ac.in}
\author[2]{\fnm{Chittaranjan} \sur{Hens}}
%\email{chittaranjanhens@gmail.com}
\author[3]{\fnm{Suman} \sur{Saha}}
%\email{suman.saha@vit.ac.in}

%% CORRESPONDING AUTHOR
\author*[4,5]{\fnm{Tanujit} \sur{Chakraborty}}
\email{tanujit.chakraborty@sorbonne.ae}

%% === AFFILIATIONS ===
\affil[1]{\orgdiv{Department of Physics}, 
          \orgname{National Institute of Technology Sikkim},
          \orgaddress{\city{Sikkim}, \postcode{737139}, \country{India}}}

\affil[2]{\orgdiv{Centre for Computational Natural Sciences and Bioinformatics}, 
          \orgname{International Institute of Information Technology},
          \orgaddress{\city{Hyderabad}, \postcode{500032}, \country{India}}}

\affil[3]{\orgdiv{School of Electronics Engineering},
          \orgname{Vellore Institute of Technology},
          \orgaddress{\city{Chennai}, \state{Tamil Nadu}, \postcode{600127}, \country{India}}}

\affil[4]{\orgdiv{SAFIR}, 
          \orgname{Sorbonne University Abu Dhabi},
          \orgaddress{\city{Abu Dhabi}, \country{UAE}}}

\affil[5]{\orgdiv{Sorbonne Center for AI}, 
          \orgname{Sorbonne University},
          \orgaddress{\city{Paris}, \country{France}}}

\abstract{

We study how the 2024 U.S. presidential election, viewed as a major political risk event, affected cryptocurrency markets by distinguishing human-driven peer-to-peer stablecoin transactions from automated algorithmic activity. Using structural break analysis, we find that human-driven Ethereum Request for Comment 20 (ERC-20) transactions shifted on November 3, two days before the election, while exchange trading volumes reacted only on Election Day. Automated smart-contract activity adjusted much later, with structural breaks appearing in January 2025. We validate these shifts using surrogate-based robustness tests. Complementary energy-spectrum analysis of Bitcoin and Ethereum identifies pronounced post-election turbulence, and a structural vector autoregression confirms a regime shift in stablecoin dynamics. Overall, human-driven stablecoin flows act as early-warning indicators of political stress, preceding both exchange behavior and algorithmic responses.}

\keywords{Political uncertainty, Cryptocurrency markets, Stablecoins, Human–Algorithmic divergence, Structural breaks, Early warning system}

\maketitle

\section{Introduction}
\label{Introduction}

Exogenous events such as political elections, geopolitical conflicts, pandemics, and macroeconomic policy announcements are widely recognized as powerful forces that disrupt complex systems, including stock markets, commodity markets, and global economies~\cite{zhang2009estimating,cutler1988moves,beber2010cannot,corbet2020impact,mukhia2024complex}. Financial markets often react sharply to such events, reflecting changes in investor sentiment, expectations, and perceptions of risk. Theoretical frameworks establish that markets are frequently dominated by political discourse, where news about government actions and electoral prospects significantly influences asset prices~\cite{pastor2013political}. These studies highlight that political uncertainty can generate both short-term volatility and long-term structural shifts in financial systems~\cite{zhang2009estimating,bloom2009impact}. It is observed that over 36\% of stock market movements are driven by political news as compared to 23\% from macroeconomic news and 11\% from corporate earnings~\cite{bialkowski2022high}. This evidence suggests that financial markets are not only driven by fundamentals but are highly sensitive to evolving political narratives. As financial systems become more globally integrated and information-driven, political shocks now propagate more rapidly, amplifying volatility and uncertainty across asset classes.

In recent years, the emergence of cryptocurrency markets has added a new dimension to the interaction between politics and finance. Decentralized digital assets operate outside traditional institutional boundaries, yet they increasingly exhibit sensitivity to macroeconomic and political developments~\cite{chen2020blockchain,omrane2025exploring,corbet2020cryptocurrency}. The global reach of cryptocurrencies and the speed of information flow through online and social platforms create conditions under which political narratives can immediately influence trading and investment behavior~\cite{shiller2020narrative}. Stablecoins such as Tether (USDT) and USD Coin (USDC) have become key components of the cryptocurrency market. By combining blockchain efficiency with fiat currency stability~\cite{ante2021influence}, stablecoins mitigate the volatility of major cryptocurrencies like Bitcoin (BTC) and Ethereum(ETH)~\cite{dyhrberg2016bitcoin,katsiampa2017volatility}. They play an essential role in liquidity provision, trading pairs, and market stability~\cite{lyons2023keeps,coinmarketcap2025stablecoin}. However, despite their design for stability, stablecoins remain exposed to systemic risks from external shocks, particularly political uncertainty, which can affect their pegs, transaction volumes, and user behavior.

Some of the key contributions relevant to this direction are discussed. De Blasis et al. (2023) investigate the May 2022 collapse of USDT using Baba, Engle, Kraft, and Kroner (1990) (BEKK) models, finding that design differences among stablecoins significantly influence the direction, magnitude, and duration of their responses to the crisis~\cite{de2023intelligent}. Grobys and Huynh (2022) examine whether jumps in USDT Granger-cause Bitcoin returns, finding that positive jumps in USDT significantly predict subsequent negative Bitcoin price changes, highlighting inefficiencies in the cryptocurrency market~\cite{grobys2022tether}. Y-H Lee et al. (2025) develop predictive models to assess the risk of stablecoin depegging, focusing on the top four USD-pegged stablecoins. Their findings highlight that major cryptocurrency fluctuations (BTC and ETH) significantly influence stablecoin stability, while sentiment variables add limited predictive power~\cite{lee2025stablecoin}. Gadzinski et al. (2024) analyze the co-instability of stablecoins by detecting structural breaks and spillover effects using Dynamic Time Warping and (Dynamic Conditional Correlation) DCC-GARCH models. They show that algorithmic stablecoins were most vulnerable during crashes, while fiat-backed tokens displayed greater resilience~\cite{gregory2024break}. Eichengreen et al. (2025) analyze the devaluation risk of stablecoins by constructing market-based measures of run risk from Tether spot and futures prices. They find that partial default risk can reach up to 12\% points, driven largely by Bitcoin volatility, transaction velocity, and redemption activity~\cite{eichengreen2025stablecoin}. Recently, the role of political signal quality in predicting cryptocurrency returns has been examined, focusing on the “crypto president” narrative during the 2024 U.S. presidential election. This study found that political news influences cryptocurrency prices in the short term and that sentiment-based measures outperform historical cryptocurrency data in predictive power~\cite{jabeur2025crypto}.

%Ben Jabeur et al. (2025)~\cite{jabeur2025crypto} examine the role of political signal quality in predicting cryptocurrency returns, focusing on the “crypto president” narrative during the 2024 U.S. presidential election. They find that political news influences cryptocurrency prices in the short term and that sentiment-based measures outperform historical cryptocurrency data in predictive power. Building on this evidence, our study extends the analysis by moving beyond price-based and sentiment-driven perspectives to explore how political uncertainty is transmitted through blockchain-native activity. Specifically, we investigate how political shocks affect blockchain transaction behavior, distinguishing between human-driven (EOA–EOA) and algorithmic (SC–SC) activity, and examine whether these behavioral patterns can serve as early indicators of market turbulence.

However, these studies mainly rely on exchange-based data and text sentiment analysis, leaving a gap in understanding how blockchain-based transaction patterns, especially those separating human and algorithmic activity, respond to political shocks.  Building on this evidence, our study extends the analysis by moving beyond price-based and sentiment-driven perspectives to explore how political uncertainty is transmitted through blockchain-based activity. Specifically, we investigate how political shocks affect blockchain transaction behavior, distinguishing between human-driven and algorithmic activity, and examine whether these behavioral patterns can serve as early indicators of market turbulence. In this study, we address this gap by examining the 2024 U.S. Presidential Election as a major political shock and analyzing its impact on cryptocurrency market dynamics through ERC-20 blockchain data. Our analysis captures the temporal sequence of human versus automated responses to political uncertainty, providing a new perspective on how behavioral and structural patterns develop during periods of market stress.

The contribution of our study can be summarized as follows. 
\begin{itemize}
    \item We distinguish between Externally Owned Account (EOA–EOA) transactions, which capture human-driven behavior, and Smart Contract (SC–SC) transactions, which represent automated algorithmic activity, in the context of political shocks.

    \item We employ a multi-method approach combining structural break analysis, Hilbert-Huang Transform, and structural vector autoregression to identify both the timing and transmission mechanisms of election-induced shocks.

    \item We demonstrate that human-driven blockchain activity serves as a leading indicator for subsequent market turbulence, providing an early indicator before exchange-based trading reacts and months before automated systems recalibrate. The 2024 U.S. election serves as a critical example validating the human-driven signal as a component of the early warning system for the politically induced financial stress. 
    
\end{itemize}
    
    These findings have practical implications for investors, risk managers, and policymakers navigating periods of political uncertainty in cryptocurrency markets.  For the investors, this early signal offers a clear, data-driven insight for making timely portfolio adjustments. For risk managers, it helps to improve the market prediction model by adding real-time blockchain data.  For policymakers, it provides the crucial understanding of how political stress spreads in the financial system, which is vital to make effective policies and maintaining a stable economy.

The remainder of this paper is organized as follows. Section~\ref{Data} explains the data sources and preprocessing steps. Section~\ref{Methodology} describes the methodological framework used in the analysis. Section~\ref{result} reports the main empirical results, including the structural break tests, Hilbert spectrum analysis, and structural vector autoregression. Section~\ref{conclu} concludes with the key findings and their implications for investors, policymakers, and future research.

\section{Problem definition and data  description}
\label{Data}

In this study, we analyze one year of ERC-20 blockchain transaction data, covering the period from March 2024 to February 2025~\cite{zhang2009estimating, mukhia2025universal, xblock2025eth}. This one-year timeframe was selected to capture both pre-election and post-election market dynamics surrounding the 2024 U.S. Presidential election. The ERC-20 transaction dataset contains nine columns, each providing specific information about ERC-20 token transfers. Each column is summarized in Table~\ref{tab:dataset_columns}. For our analysis, we utilize the following columns: timeStamp, tokenAddress, fromIsContract, toIsContract, and value.

\begin{table}[h]
\centering
\caption{Data description of ERC-20 blockchain transaction dataset columns and their descriptions. The dataset records detailed information for each token transfer, including the block timestamp, token contract address, sender and receiver accounts, and transfer amount.}
\label{tab:dataset_columns}
\renewcommand{\arraystretch}{1.5}

\begin{tabular}{@{}ll@{}}
\toprule
\textbf{Column} & \textbf{Description} \\
\midrule
\texttt{timeStamp} & The time when the block was created; all transactions within the same block share this timestamp. \\
\texttt{tokenAddress} & The hash value of the ERC-20 token’s contract, which serves as its unique identifier. \\
\texttt{from} & The Ethereum address of the sender of the ERC-20 token. \\
\texttt{to} & The Ethereum address of the receiver of the ERC-20 token. \\
\texttt{fromIsContract} & Indicates whether the sender is an SC (1) or an EOA (0). \\
\texttt{toIsContract} & Indicates whether the receiver is an SC (1) or an EOA (0). \\
\texttt{value} & The amount of tokens transferred, in the token's base unit. \\
\botrule
\end{tabular}

\end{table}

We first pre-processed the ERC-20 blockchain transaction dataset before conducting the analysis. On the Ethereum blockchain, there are thousands of ERC-20 tokens, including both stablecoins and non-stable assets. Among these, we focused on the two largest stablecoins by market share, USDT and USDC, which together account for more than 80\% of the total stablecoin market capitalization~\cite{coinmarketcap2025stablecoin}. Each token was identified by its verified contract address on Etherscan~\cite{etherscan2025eth}:

\begin{itemize}
\item USDT: 0xdac17f958d2ee523a2206206994597c13d831ec7
\item USDC: 0xa0b86991c6218b36c1d19d4a2e9eb0ce3606eb48
\end{itemize}

In the ERC-20 framework, two types of accounts exist. EOAs are controlled by individuals through public-private key pairs and represent human users~\cite{chen2020traveling,mukhia2025universal}. Whereas SCs are governed by executable code stored within the blockchain account itself~\cite{szabo1997idea,kolvart2016smart}. These account types are indicated in the dataset by the columns fromIsContract and toIsContract, summarized in Table~\ref{tab:dataset_columns}. A transaction is labeled as EOA–EOA (human-driven) when fromIsContract = 0 and toIsContract = 0, while SC–SC transactions are identified when fromIsContract = 1 and toIsContract = 1. Each transaction in the ERC-20 data set is recorded with a timestamp in UTC format~\cite{wikipedia2025unixtime}. For example, the value 1715385600 corresponds to 2024-11-05, based on the Unix epoch starting from 1970-01-01 00:00:00 UTC. In this study, we converted these UTC timestamps into a human-readable format (YYYY-MM-DD) for analysis. To capture how human-driven and automated execution responded to the U.S. Presidential Election shock, we focus on two transaction categories: EOA–EOA and SC–SC. Table~\ref{tab:transaction_summary} summarizes the distribution of these transaction types.

\begin{table}[h]
\centering
\caption{Summary of ERC-20 transaction types based on sender and receiver account classification. Transactions are divided into human-driven (EOA--EOA) and automated (SC--SC) categories using the fromIsContract and toIsContract. This classification helps us to distinguish between human-initiated transfers and code-executed activities on the Ethereum blockchain, providing a foundation for analyzing behavioral and algorithmic responses to political shocks.}
\label{tab:transaction_summary}
\renewcommand{\arraystretch}{1.5}

\begin{tabular}{@{}llll@{}}
\toprule
\textbf{fromIsContract} & \textbf{toIsContract} & \textbf{Transaction Type} & \textbf{Number of Transactions} \\
\midrule
0 & 0 & EOA--EOA (Human-driven) & 50,597,281 \\
1 & 1 & SC--SC (Automated code-driven) & 24,393,131 \\
\midrule
\multicolumn{3}{r}{Total Transactions} & 102,575,812 \\
\botrule
\end{tabular}

\end{table}

On the Ethereum blockchain, the value of USDT and USDC is stored in their smallest indivisible unit, not directly in U.S. dollars (USD). Both tokens use six decimal places, which means that the raw number in the value column is the actual amount multiplied by $10^{6}$. To convert this into the real transaction amount in USD, we divide the raw value by $10^{6}$:

\begin{equation}
\text{USD Value} = \frac{\texttt{value}}{10^{6}}
\end{equation}

For example, a recorded value of 1{,}000{,}000 corresponds to 1.0 USD. Duplicate, failed, and zero-value transactions were removed to reduce noise. Finally, daily transaction volumes were aggregated separately for EOA–EOA and SC–SC activities, forming two time series that represent human-driven and automated blockchain behavior. These time series were merged with daily trading volume data for USDT and USDC, as well as closing prices for BTC and Ethereum ETH, to align blockchain activity with broader market conditions around the 2024 US presidential election. This allows us to correlate changes in blockchain transaction behavior with shifts in market-wide liquidity, trading activity, and the prices of major digital assets, particularly around the 2024 U.S. Presidential election.

\section{Methodology}
\label{Methodology}

To study how the 2024 U.S. Presidential election affected cryptocurrency market dynamics, we applied several complementary analytical methods. We employed several statistical tests, such as the Bai-Perron test, to identify significant regime changes in stablecoin ERC-20 blockchain transactions and trading volumes, and the Augmented Dickey-Fuller test for stationarity check. We then applied the Hilbert–Huang Transform to detect extreme events and quantify market turbulence in major cryptocurrencies. Next, we implemented Amplitude-Adjusted Fourier Transform surrogate testing to validate the statistical significance of the identified breakpoints against noise. Finally, we applied structural vector autoregression with Wald tests to analyze volatility spillovers and confirm structural changes between pre- and post-election periods. Further details on each of these techniques are discussed below.

\subsection{Statistical tests}

In this section, we employ various statistical tests to determine key time series properties. The Augmented Dickey-Fuller test assesses stationarity, the Bai-Perron test detects structural breaks, which are sudden shifts in the time series mean, with the SupF test validating their statistical significance, as presented below.

\subsubsection {Augmented Dickey-Fuller test}
\label{ADF test}

The Augmented Dickey–Fuller (ADF) test is a widely used statistical test for examining the stationarity of time series data \cite{amarasinghe2015dynamic,paramati2011empirical,yang2020novel}. This test is the extended form of the original Dickey–Fuller (DF) test, which models the time series as an autoregressive process of order one, AR(1)~\cite{dickey1981likelihood}. To address the problem of autocorrelation in the DF test, Said and Dickey introduced the ADF test by incorporating additional lagged terms of the dependent variable into the regression \cite{said1984testing,herranz2017unit}.

Let's say $y_t$ follows an autoregressive process. The ADF regression can be written as
\begin{equation}
    \Delta y_t = \beta + \gamma t + \alpha y_{t-1} + \sum_{j=1}^{n} \psi_j \Delta y_{t-j} + \varepsilon_t,
    \label{ADF_least}
\end{equation}
where $\Delta$ is the first-difference operator, $\beta$ is a constant, $\gamma$ captures a deterministic trend, and $\varepsilon_t$ is white noise.

The ADF test is used to determine whether a time series contains a unit root, which would suggest that the series is non-stationary. In this model, the null hypothesis assumes the presence of a unit root in the series, while the alternative hypothesis assumes that the time series is stationary and does not have a unit root \cite{paparoditis2018asymptotic}. The hypotheses for the ADF test are

\begin{equation}
H_{0}: \alpha = 1 \;\;\Rightarrow\;\; y_{t} \sim I(1)
\end{equation}

\begin{equation}
H_{1}: |\alpha| < 1 \;\;\Rightarrow\;\; y_{t} \sim I(0)
\end{equation}

The test statistic is based on the least squares estimate of Eq.~\ref{ADF_least}, expressed as  

\begin{equation}
t_{\alpha=1} = \frac{\hat{\alpha} - 1}{SE(\alpha)}
\end{equation}

where $\hat{\alpha}$ is the least squares estimate and $SE(\alpha)$ denotes its standard error.

The ADF test reports the $t$-statistic, $p$-value, and critical values at different significance levels. Since the distribution of $t_\alpha$ under the null is non-standard, critical values have been provided by DF (1979) and refined by MacKinnon (1996)~\cite{mackinnon1996numerical}. If the computed statistic is smaller than the critical value, the null hypothesis of a unit root is rejected, and the series is considered stationary. Otherwise, the series is deemed non-stationary and must be differenced to achieve stationarity~\cite{kim2017unit}. In our study, we used the ADF test to verify the stationarity conditions required before estimating the structural vector autoregression model and conducting Hilbert Spectrum extreme event analysis.
%Next, we test for structural breaks in the time series with the Bai-Perron methodology, as presented in detail below.

\subsubsection{Bai-Perron test}
\label{sec:structural_break_analysis}

Financial time series often exhibit sudden shifts in their statistical properties due to various economic, political, or market-related events~\cite{sengupta2025forecasting,besher5705079modeling}. Examples include financial crises, policy interventions, elections, and global shocks, all of which can lead to changes in the underlying data-generating process~\cite{mahata2020identification,chaudhuri2003random,worthington2007gold}. Identifying such structural changes is crucial for understanding market dynamics, evaluating risk, and accurately modeling financial behavior~\cite{chakraborty2025neural}.

To detect these shifts, we employ the multiple structural break methodology proposed by Bai and Perron (BP)~\cite{bai1998estimating,bai2003computation}, which provides an endogenous framework for identifying unknown breakpoints in time series without requiring prior knowledge of their locations. The BP test jointly estimates the number and timing of breaks as well as the model parameters associated with each regime by minimizing the global sum of squared residuals (SSR).

Consider a multiple structural change regression model with $m$ breakpoints
\begin{equation}
    y_t = x_t^{\prime}\beta + z_t^{\prime}\delta_j + u_t, 
    \quad t = T_{j-1}+1,\dots,T_j, \quad j = 1,\dots,m+1,
\end{equation}
where $y_t$ is the observed dependent variable, $x_t \in \mathbb{R}^p$ and $z_t \in \mathbb{R}^q$ are vectors of covariates, $\beta$ and $\delta_j$ are coefficient vectors, and $u_t$ is the error term. . The index $j$ identifies the different regimes in the time series, with each regime $j = 1,\dots,m+1$ covering the observations between two successive break dates. The break dates $(T_1,\dots,T_m)$ are unknown and estimated directly from the data.

For a given partition $(T_1,\dots,T_m)$, the least-squares estimators $\hat{\beta}$ and $\hat{\delta}_j$ are obtained by minimizing the sum of squared residuals(SSR)
\begin{equation}
    \text{SSR}(T_1,\dots,T_m) = \sum_{i=1}^{m+1} 
    \sum_{t=T_{i-1}+1}^{T_i} 
    \left(y_t - x_t^{\prime}\beta - z_t^{\prime}\delta_i\right)^2.
\end{equation}
The optimal break dates are then given by
\begin{equation}
    \{\hat{T}_1,\dots,\hat{T}_m\} 
    = \arg \min_{(T_1,\dots,T_m)} \text{SSR}(T_1,\dots,T_m),
\end{equation}
where the minimization is subject to a trimming parameter to ensure sufficient observations within each regime.

A dynamic programming algorithm~\cite{bai2003computation} is used to evaluate all possible partitions efficiently and select the global minimum of the SSR. This approach provides consistent estimates of both the break dates and the associated regime-specific model parameters.

In this study, we use log-transformed data and model it as a mean-shift process
\begin{equation}
y_t = \mu_j + u_t,
\quad t = T_{j-1}+1,\dots,T_j, \quad j = 1,\dots,m+1,
\end{equation}
where $y_t$ is the log-value at time $t$, $\mu_j$ is the regime-specific mean, 
$u_t$ is the error term, $T_j$ are the breakpoints, and $m$ is the number of 
estimated breaks. A breakpoint indicates a shift from one mean level $\mu_j$ 
to another $\mu_{j+1}$, capturing structural changes in the series.

%To validate whether the breakpoints identified by the Bai-Perron procedure represent statistically significant structural changes rather than random fluctuations, we implement a series of formal hypothesis tests. These tests provide complementary approaches to verify parameter stability and confirm the significance of detected regime shifts.

To validate whether the breakpoints identified by the BP procedure represent statistically significant structural changes rather than random fluctuations, we implement the SupF hypothesis tests. This test verifies parameter stability and confirms the significance of detected regime shifts.

\subsubsection{SupF test}
\label{sec:supf_test}

The SupF test~\cite{bai1998estimating,bai2003critical} is used to examine whether at least one structural break exists within the time series. The hypotheses for the SupF test can be stated as
\begin{align}
H_0 &: \text{No structural break exists}, \\
H_1 &: \text{At least one structural break exists}.
\label{eq:hypothesis}
\end{align}

The SupF statistic is computed as
\begin{equation}
\text{SupF} = \max_{\tau \in \Lambda} F(\tau),
\end{equation}
where $F(\tau)$ is the F-statistic testing parameter stability at break fraction $\tau$, and $\Lambda = [\tau_{\text{min}}, \tau_{\text{max}}]$ represents the set of all possible break fractions within the trimmed time series.

The F-statistic for a candidate break fraction $\tau$ is calculated as
\begin{equation}
F(\tau) = \frac{(SSR_0 - SSR_1)/q}{SSR_1/(T - 2q)},
\end{equation}
where $SSR_0$ is the sum of squared residuals under $H_0$, $SSR_1$ is the sum of squared residuals under $H_1$, $q$ is the number of parameters subject to change, and $T$ is the total number of observations.

We reject $H_0$ at the 5\% significance level if the p-value associated with the SupF statistic is less than 0.05, concluding that at least one structural break is present. In our mean-shift model, this corresponds to a statistically significant change in the average level of the log-transformed series. We used the BP test together with the SupF test to identify and validate structural shifts in stablecoin and cryptocurrency markets around the 2024 U.S. election.

\subsection{Transformation methods}

After establishing the presence of structural breaks, this section presents the methodologies to detect, validate, and analyze these breaks. With the Hilbert spectrum, we detect the timing of the sudden change. The Amplitude-Adjusted Fourier Transform then provides the statistical validation of the BP test and Hilbert spectrum findings. Subsequently, structural vector autoregression analyzes the dynamic relationships between the time series across regimes. Finally, the Wald test is applied to the pre- and post-breakpoint regimes of the structural vector autoregression analyses to quantify the significance of the structural change, as presented in detail in the following subsections. 

\subsubsection{Hilbert–Huang Transform (HHT)}
\label{Hilbert spectrum}

The Empirical Mode Decomposition (EMD) technique is widely applied to decompose nonlinear and nonstationary time series data by separating the signal into a set of Intrinsic Mode Functions (IMFs). Each IMF is associated with a distinct time scale \cite{huang1998empirical,huang2003applications,mahata2020identification}.

A time series signal component is considered an IMF when the following two conditions are satisfied 

\begin{itemize}
    \item The number of extrema and zero-crossings should be the same, or differ by one.
    \item The mean of the envelopes formed by the local maxima and minima should be zero.

\end{itemize}

The steps for decomposing a time series signal into IMFs are discussed below.

\begin{enumerate}
    \item For a input time series dataset $D_t$,  construct the upper envelope ($UE_t$) from its local maxima and the lower envelope ($LE_t$) from its local minima using spline fitting.
   \item Calculate the mean of the upper and lower envelopes,
   
    \begin{equation}
        m_t = \frac{UE_t + LE_t}{2},
    \end{equation}
    and subtract this mean from the original time series signal to obtain the updated time series.
    
 \begin{equation}
        ND_t = D_t - m_t.
      \end{equation}

    \item Repeat the process on $ND_t$ until it satisfies the two IMF conditions.  When these conditions are satisfied, the resulting time series is considered as the first IMF, denoted by $IMF_1$.

    \item For the next IMF, calculate the residual
    \begin{equation}
        N_t = D_t - IMF_1,
      \end{equation}

    and apply the same step to $N_t$.
    
    \item The decomposition process continues until the final residual becomes monotonic, which is consider to as the residue. The residue show the overall trend of the original dataset.
    \item The original dataset can then be reconstructed as the sum of all IMFs and the residue
    \begin{equation}
        D_t = \sum_{j=1}^{n} IMF_j + \text{residue},
     \end{equation}

    where $n$ is the total number of IMFs.
\end{enumerate}

For each IMF, the instantaneous frequency $\omega$ is obtained by applying the Hilbert transform, defined as

\begin{equation}
    H(t) = \dfrac{1}{\pi} \, \text{P.V.} \int_{-\infty}^{\infty} \dfrac{\text{IMF}(\tau)}{t - \tau} \, d\tau,
    \label{eq:hilbert}
\end{equation}
where P.V. represents the Cauchy principal value. The instantaneous phase $\phi(t)$ is expressed as~\cite{huang1998empirical}

\begin{equation}
    \phi(t) = \tan^{-1} \left( \dfrac{H(t)}{\text{IMF}(t)} \right),
    \label{eq:phase}
\end{equation}

and the instantaneous frequency $\omega$ is obtained from

\begin{equation}
    \omega = \dfrac{d\phi}{dt}.
    \label{eq:freq}
\end{equation}

The Hilbert spectrum $H(t,\omega)$ (HS), which represents the time-frequency distribution of the  time series signal, is calculated using

\begin{equation}
    H(t,\omega) = \Re \left\{ \sum_{i} K_i(t) \, e^{j \int \omega(t) \, dt} \right\},
    \label{eq:spectrum}
\end{equation}
where $K_i(t)$ is the instantaneous amplitude and $\Re\{\cdot\}$ denotes the real part.

The instantaneous energy, denoted as $IE(t)$, can be calculated from the Hilbert spectrum and is defined as

\begin{equation}
    IE(t) = \int_{\omega} H^2(t,\omega) \, d\omega.
    \label{eq:ie}
\end{equation}

The normalized instantaneous energy $IE_N(t)$ is defined as

\begin{equation}
    IE_{N}(t) = \dfrac{IE(t)}{\max\left[IE(t)\right]}.
    \label{eq:ien}
\end{equation}

An event is classified as an EE if $IE_{N}(t)$ exceeds a threshold value. The threshold energy $E_{th}$ is determined by

\begin{equation}
    E_{th} = E_{\mu} + B\sigma,
    \label{eq:eth}
\end{equation}
where $E_{\mu}$ is the average energy, $\sigma$ is the standard deviation, and $B$ is a constant parameter. For the present analysis, we set $B = 4$. To ensure that the detected points are not just random fluctuations, we present the Amplitude-Adjusted Fourier Transform to provide us with the statistical validation.

In this study, we applied Eq.~\eqref{eq:spectrum} to compute $H(t,\omega)$ for the combined IMFs. By locating the regions of maximum energy concentration in the $H(t,\omega)$ spectrum, the occurrence time of an Extreme Event (EE) or sudden price change can be identified.  
To quantify an EE, we evaluate the instantaneous energy of the combined IMFs using Eq.~\eqref{eq:ie}. In our case, this method allows us to pinpoint extreme post-election movements in BTC and ETH and link them to stress originating from stablecoin activity.

%\subsubsection{}
\subsubsection{Amplitude-Adjusted Fourier Transform (AAFT)}
\label{method:AAFT}

Surrogate data methods provide a statistical framework for testing whether observed structures in a time series reflect genuine nonlinear dynamics or arise from random linear processes \cite{schreiber2000surrogate,lancaster2018surrogate}. Among the common algorithms for surrogate generation, the Fourier Transform (FT) method represents the earliest restricted implementation \cite{theiler1992testing}. The FT algorithm proceeds as follows.

Given original time series data \(x(t)\), we perform the Fourier transformation to obtain
\begin{equation}
   X(f) = A(f)e^{i\phi(f)}
\end{equation}
where \(A(f)\) is the amplitude and \(\phi(f)\) is the phase. We then multiply by \(e^{i\varphi(f)}\) to obtain
\begin{equation}
   \tilde{X}(f) = A(f)e^{i[\phi(f) + \varphi(f)]} 
\end{equation}
where \(\varphi(f)\) is a random phase obeying independent uniform distribution on \([0,2\pi]\). Finally, the Fourier inverse transformation gives
\begin{equation}
    \tilde{x}(t) = \mathcal{F}^{-1}\{\tilde{X}(f)\} = \mathcal{F}^{-1}\{X(f)e^{i\varphi(f)}\}
\end{equation}
The resulting \(\tilde{x}(t)\) represents the surrogate data generated by the FT algorithm.

However, standard FT surrogates assume Gaussian distributions and fail to preserve the amplitude distribution of non-Gaussian real-world data. The Amplitude-Adjusted Fourier Transform (AAFT) method extends the FT approach to preserve both the power spectrum and the amplitude distribution \cite{theiler1992testing,kugiumtzis1999test}. The AAFT algorithm proceeds through the following steps.

\begin{enumerate}
\item Generate Gaussian white noise \(n(t)\) with the same ordering (ranks) as the original data \(x(t)\).
\item Apply the FT algorithm to generate surrogate data \(\tilde{n}(t)\) from the Gaussian white noise.
\item Reorder the original data \(x(t)\) according to the rank order of \(\tilde{n}(t)\) to obtain the final surrogate data \(\tilde{x}(t)\).
\end{enumerate}

The null hypothesis assumes that the observed time series is generated by a rescaled linear Gaussian process, i.e., a stationary linear Gaussian system passed through an invertible measurement function.

\begin{equation}
    s_n = h(x_n), \quad x_n = \sum_{i=1}^{M} a_i x_{n-i} + \sum_{i=0}^{N} b_i \eta_{n-i},
\end{equation}
where \(h(\cdot)\) is the invertible measurement function, and \(\eta_n\) represents Gaussian white noise.

The AAFT method ensures that the surrogate datasets preserve the linear autocorrelation structure through the Fourier spectrum and retain the original amplitude distribution through rank-based remapping. This dual preservation makes AAFT particularly suitable for testing nonlinearity in financial and economic time series, which often exhibit non-Gaussian characteristics such as heavy tails and skewness.

In this study, we employ AAFT surrogate analysis with 1000 iterations to verify that the structural breaks identified by the BP test and the extreme events detected through the HS are not the result of noise or autocorrelation. This ensures that the observed changes reflect genuine structural shifts in the time series. Once the structural breaks are confirmed, we then proceed with structural vector autoregression to model their economic impact.

\subsection{Structural Vector Autoregression (SVAR)}

Structural vector autoregressive (SVAR) models are fundamental tools for empirical analysis in macroeconomics, finance, and related fields~\cite{kilian2017structural,lutkepohl2013introduction, sims1980macroeconomics}. Originally developed by Sims (1980)~\cite{sims1980macroeconomics}, vector autoregressive (VAR) models have become a standard tool for analyzing dynamic relationships among economic variables and identifying structural shocks~\cite{stock2016dynamic}. In this study, we employ an SVAR framework to identify contemporaneous causal relationships and volatility spillovers between USDT and USDC stablecoin flows, with particular focus on changes induced by the 2024 U.S. Presidential election.

Let \(Y_t = [y_{1,t}, y_{2,t}]'\) denote a 2-dimensional vector of stationary time series representing daily stablecoin transaction volumes. The reduced-form VAR(\(p\)) model is written as

\begin{equation}
Y_t = c + \sum_{i=1}^{p} \Phi_i Y_{t-i} + u_t,
\end{equation}

where, \(c\) is a vector of constants, \(\Phi_i\) are \(2 \times 2\) coefficient matrices for lag \(i\) and \(u_t\) is a vector of serially uncorrelated reduced-form innovations with covariance matrix \(\Sigma_u = \mathbb{E}[u_t u_t']\).

\subsubsection{Identification of structural break}

The reduced-form innovations \(u_t\) are linear combinations of mutually orthogonal structural shocks \(\varepsilon_t\), which satisfy \(\mathbb{E}[\varepsilon_t \varepsilon_t'] = I\). The structural form is

\begin{equation}
A_0 u_t = \varepsilon_t \quad \Rightarrow \quad u_t = A_0^{-1} \varepsilon_t,
\end{equation}

which implies the covariance relationship

\begin{equation}
\Sigma_u = A_0^{-1} (A_0^{-1})'.
\end{equation}

To achieve identification, we impose a recursive (Cholesky) structure by restricting \(A_0^{-1}\) to be lower triangular~\cite{bernanke2005measuring}. This allows unique recovery of the structural impact matrix via Cholesky decomposition

\begin{equation}
\Sigma_u = PP' \quad \Rightarrow \quad \widehat{A_0^{-1}} = P,
\end{equation}

where \(P\) is the lower-triangular Cholesky factor. The variable ordering \(Y_t = [\text{USDC}_t, \text{USDT}_t]'\) reflects the identifying assumption that structural shocks to USDC can affect USDT contemporaneously, while shocks to USDT affect USDC only with a one-period lag.

We address the non-stationarity of financial time series by applying a logarithmic transformation with first-differencing to USDC and USDT volumes. This dual transformation converts the series to stationary percentage changes while handling zero volumes through the log(1+x) formulation. For lag selection, we employ the Akaike Information Criterion (AIC). This approach ensures we capture meaningful dynamics without overfitting, providing a balanced representation of the temporal relationships between stablecoin flows. In our analysis, the SVAR model enables us to examine how shock transmission between USDC and USDT changes across regimes, providing evidence that the 2024 U.S. election produced a structural shift in their contemporaneous interactions.

\subsubsection{Wald test}

To formally test for election-induced structural breaks, we implement a Wald test comparing parameter estimates obtained from the pre-election and post-election~\cite{wald1943tests}. Let $\hat{\theta}_{\text{pre}}$ and $\hat{\theta}_{\text{post}}$ denote the vectors of estimated model parameters from the two periods. The parameter vector $\theta$ consists of the VAR autoregressive coefficients, which capture the lagged interaction dynamics between USDC and USDT. Accordingly, $\hat{\theta}_{\text{pre}}$ and $\hat{\theta}_{\text{post}}$ represent the stacked lag coefficients estimated separately for the pre-election and post-election periods.

We test the one-sided hypothesis that election-induced uncertainty increases volatility and spillover strength, which corresponds to an increase in the magnitude of the dynamic coefficients:

\begin{equation}
H_0: \hat{\theta}_{\text{post}} \leq \hat{\theta}_{\text{pre}}, 
\qquad 
H_1: \hat{\theta}_{\text{post}} > \hat{\theta}_{\text{pre}}.
\end{equation}

The Wald statistic for testing this hypothesis is given by
\begin{equation}
W = (\hat{\theta}_{\text{post}} - \hat{\theta}_{\text{pre}})' 
\left( \widehat{\Sigma}_{\text{pre}} + \widehat{\Sigma}_{\text{post}} \right)^{-1} 
(\hat{\theta}_{\text{post}} - \hat{\theta}_{\text{pre}}),
\end{equation}
where $\widehat{\Sigma}_{\text{pre}}$ and $\widehat{\Sigma}_{\text{post}}$ denote the covariance matrices of the estimated parameters from the two time series subsamples. Under the null hypothesis of no structural change, the Wald statistic follows a $\chi^2$ distribution with degrees of freedom equal to the number of parameters being tested. A statistically significant Wald statistic leads to rejection of $H_0$, indicating that the 2024 U.S. Presidential election induced a structural shift in the dynamic relationships between USDC and USDT transaction flows.

\section{Results and Discussion}

\label{result}
In this section, we discuss the results of our analysis examining how the 2024 U.S. Presidential election affected cryptocurrency market behavior through multiple analytical frameworks. Existing studies mainly depend on exchange-based prices and sentiment, leaving the transmission of political uncertainty through blockchain-based activity largely unexplored. To address this gap, we distinguish between human-driven (EOA–EOA) and automated (SC–SC) transactions on the Ethereum blockchain. Our findings show a clear temporal pattern. Human activity responds first in anticipation of election-related uncertainty, while algorithmic systems react later as market conditions evolve. This pattern indicates that human-driven blockchain behavior acts an early signal of emerging market turbulence, offering insights not captured by traditional market data.

\subsection{Stationarity test}
\label{Result:ADF}

We conducted stationarity tests on the ERC-20 blockchain transaction and the trading data of USDT and USDC, along with the trading data of BTC and ETH, using the ADF test as described in Section~\ref{ADF test}. For a time series to be stationary, the test statistic should be less than the 5\% critical value, and the $p$-value should fall below the 5\% significance level, thereby rejecting the null hypothesis of a unit root. If the test statistic exceeds the 5\% critical value and the $p$-value is greater than the significance threshold, the null hypothesis cannot be rejected, indicating that the time series is non-stationary. In this study, the ADF test was applied to verify the preconditions required for SVAR and HS analyses. Table~\ref{tab:adf_stationarity_all} presents the ADF test statistics, p-values, and 5\% critical values for blockchain and trading datasets. The results show that all log-transformed series are non-stationary in levels but become stationary after first differencing. This confirms that they are integrated in order one, $I(1)$. The differenced series of USDT and USDC are therefore employed in the SVAR estimation, while for BTC and ETH, the non-stationary log-transformed trading series are retained in their original form, as HS analysis can directly accommodate non-stationary processes. This methodology ensures that the SVAR model operates on stationary time series, allowing us to accurately assess how election-induced shocks alter the regime dynamics between USDT and USDC, while the HS method utilizes the non-stationary time series of BTC and ETH prices to detect extreme events without violating its modelling assumptions. This framework provides a consistent basis for linking statistical changes in the data to real market behavior, illustrating how the 2024 U.S. election contributed to heightened volatility and accelerated spillovers across stablecoin flows and major cryptocurrency markets.

\begin{table}[h!]
\centering
\caption{The table reports Augmented Dickey–Fuller test statistics, p-values, and 5\% critical values for blockchain and trading datasets. The null hypothesis of a unit root is rejected when the test statistic is less than the 5\% critical value. All log-transformed series are non-stationary in levels but become stationary after first differencing, confirming that the series are integrated of order one, $I(1)$.}
\label{tab:adf_stationarity_all}
\renewcommand{\arraystretch}{1.3}

\begin{tabular}{@{}p{1.5cm}p{1.5cm}p{2cm}p{1.5cm}p{1.5cm}p{2cm}p{2.2cm}@{}}
\toprule
\textbf{Category} & \textbf{Asset} & \textbf{Transformation} & \textbf{Test Stat.} & \textbf{p-value} & \textbf{Critical (5\%)} & \textbf{Result} \\
\midrule
\multicolumn{7}{c}{\textbf{Blockchain Data: EOA--EOA}} \\
\midrule
USDT & EOA--EOA & Level & -1.7760 & 0.3925 & -2.864  & Non-stationary \\
     &          & First Difference & -5.9891 & 0.0000 & -2.864 & Stationary\\
USDC & EOA--EOA & Level & -1.6344 & 0.4652 & -2.864   & Non-stationary\\
     &          & First Difference & -8.6408 & 0.0000 & -2.864 & Stationary\\
\midrule
\multicolumn{7}{c}{\textbf{Blockchain Data: SC--SC}} \\
\midrule
USDT & SC--SC & Level & -1.9222 & 0.3217 & -2.864  & Non-stationary \\
     &        & First Difference & -6.4756 & 0.0000 & -2.864 & Stationary \\
USDC & SC--SC & Level & -0.3638 & 0.9160 & -2.864  & Non-stationary \\
     &        & First Difference & -7.2497 & 0.0000 & -2.864 & Stationary \\
\midrule
\multicolumn{7}{c}{\textbf{Trading Data (Exchange)}} \\
\midrule
USDT-USD & Volume & Level & -1.9571 & 0.3057 & -2.864  & Non-stationary \\
         &        & First Difference & -5.8145 & 0.0000 & -2.864 & Stationary \\
USDC-USD & Volume & Level & -2.6411 & 0.0848 & -2.864  & Non-stationary \\
         &        & First Difference & -6.4122 & 0.0000 & -2.864 & Stationary \\
BTC-USD  & Close  & Level & -1.2957 & 0.6312 & -2.864  & Non-stationary \\
         &        & First Difference & -20.2909 & 0.0000 & -2.864 & Stationary \\
ETH-USD  & Close  & Level & -1.8143 & 0.3734 & -2.864 & Non-stationary  \\
         &        & First Difference & -19.3631 & 0.0000 & -2.864  & Stationary \\
\botrule
\end{tabular}

\end{table}

\subsection{Structural Break Analysis (SBA)}
\label{Result:SBA}

In this section, we employ the Bai-Perron test~\cite{bai1998estimating,bai2003computation} to detect the break dates in the time series. To ensure the identified breakpoints represent genuine structural changes rather than random fluctuations in autocorrelated time series, we complement the BP test with AAFT surrogate analysis. This rigorous validation method preserves both the linear autocorrelation structure and amplitude distribution of the original time series while randomizing phase information, providing a robust null hypothesis against which to test statistical significance.

The purpose of this test is to examine whether the 2024 U.S. Presidential election triggered structural changes in ERC-20 blockchain transaction data and in the trading data of two major stablecoins: USDT and USDC, as well as two major cryptocurrencies, BTC and ETH. We apply the BP test across three categories of transactions:

\begin{itemize}[itemsep= 1pt, topsep=1.5pt]
\item ERC-20 blockchain transactions between EOA–EOA.
\item ERC-20 blockchain transactions between SC–SC.
\item Trading data:Stablecoins: USDT and USDC and Cryptocurrencies: BTC and ETH.
\end{itemize}

\begin{figure}[H]
   % \centering
   \includegraphics[width=0.9\linewidth, height=0.9\linewidth]{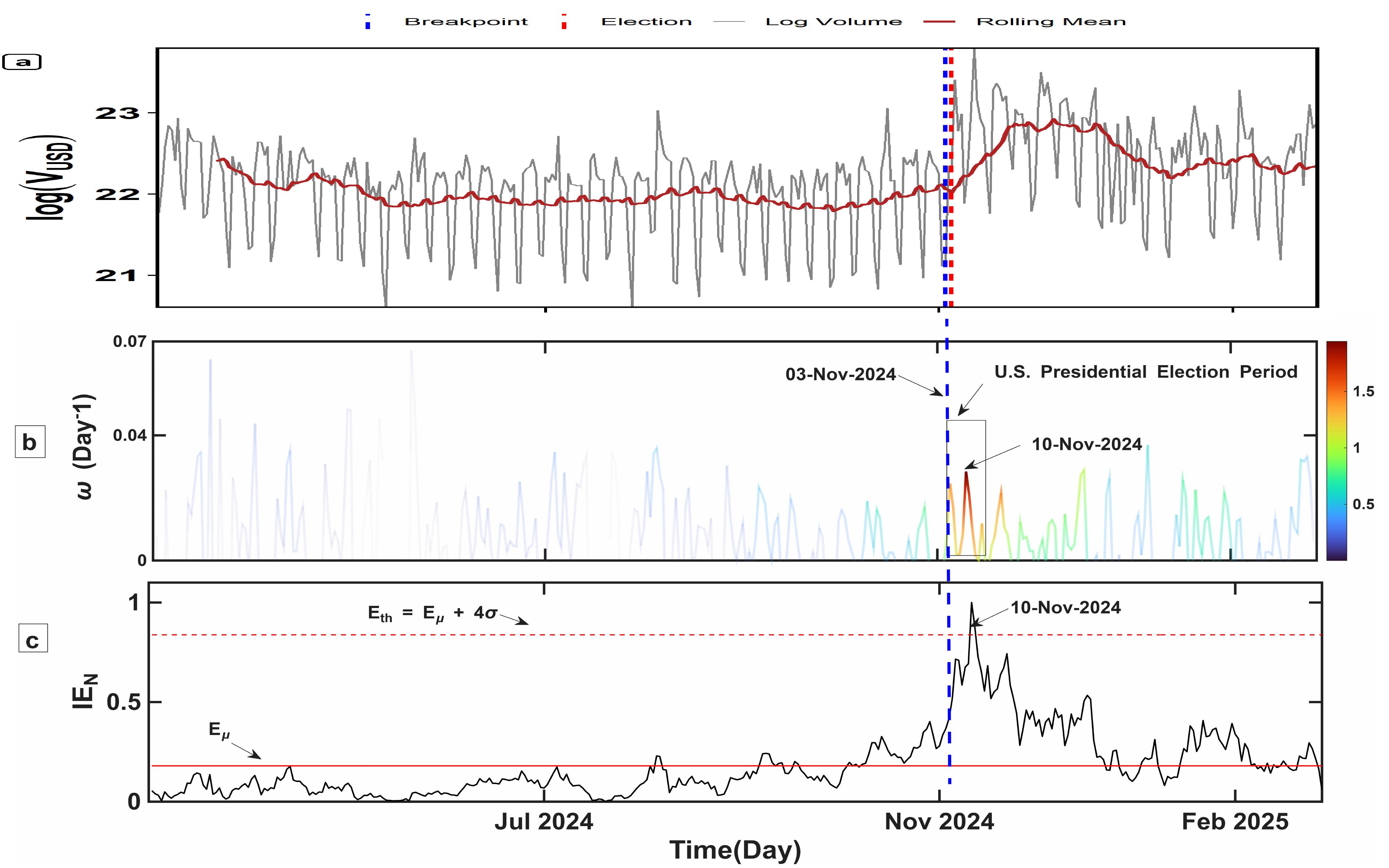}
   \centering
   \caption{
The figure illustrates the structural breakpoint, Hilbert spectrum, and instantaneous energy for blockchain USDT BTC trading data.  \textbf{(a)} Logarithm of the daily trading volume in USD, $\log(V_\mathrm{USD})$. The brown line represents the rolling mean of the trading volume, the red dashed line indicates the U.S. Presidential election Day 2024, and the blue dashed line marks the structural breakpoint.  \textbf{(b)} Hilbert Spectrum for BTC, showing the instantaneous frequency ($\omega$) over time. A period of high energy is observed during the U.S. Presidential election.   \textbf{(c)} Normalized instantaneous energy for BTC. A sudden energy surge occurs after the election period begins, peaking on November 10, 2024. This peak exceeds the statistical threshold energy level, defined as $E_\mathrm{th} = E_\mu + 4\sigma$.}
    \label{fig:USDT_HS_BTC}
\end{figure}

\begin{figure}[H]
\centering
\includegraphics[width=1.0\linewidth, height=1.2\linewidth]{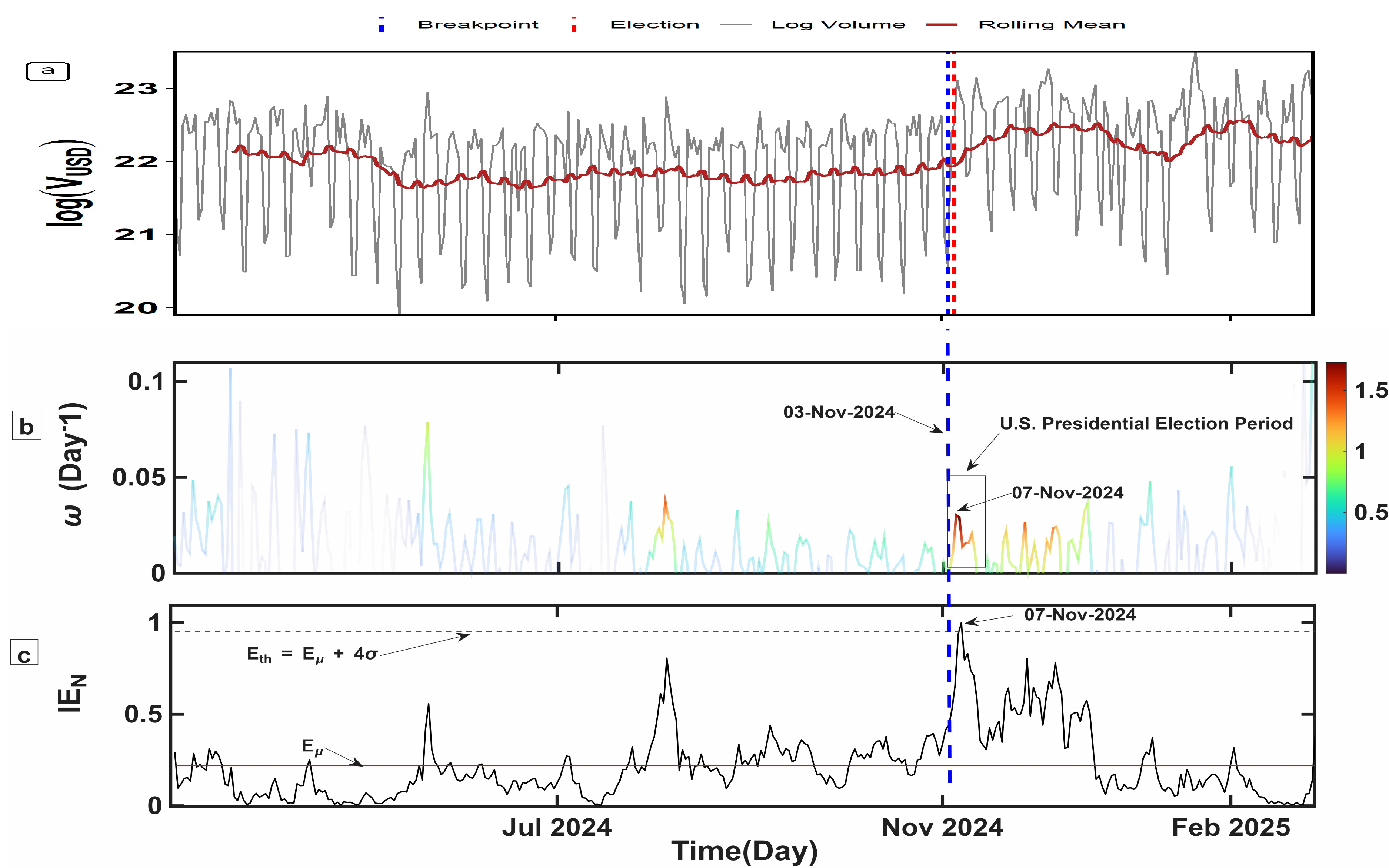} %, height=1.15
\caption{
The figure illustrates the structural breakpoint, Hilbert spectrum, and instantaneous energy for blockchain USDC and ETH trading data. \textbf{(a)} Logarithm of the daily trading volume in USD, $\log(V_\mathrm{USD})$. The brown line represents the rolling mean of the trading volume, the red dashed line indicates the U.S. Presidential election Day 2024, and the blue dashed line marks the structural breakpoint.  \textbf{(b)} Hilbert Spectrum for ETH, showing the instantaneous frequency ($\omega$) over time. A period of high energy is observed during the U.S. Presidential election.  \textbf{(c)} Normalized instantaneous energy for ETH. A sudden energy surge occurs after the election period begins, peaking on November 7, 2024. This peak exceeds the statistical threshold energy level, defined as $E_\mathrm{th} = E_\mu + 4\sigma$. 
}
\label{fig:USDC_HS_ETH}
\end{figure}

Figures~\ref{fig:USDT_HS_BTC}(a), \ref{fig:USDC_HS_ETH}(a), \ref{fig:SBA_Trading}, and \ref{fig:SBA_SC-SC} illustrate the SBA results, with structural breakpoints identified using the BP test. The brown line in these figures represents the 20-day rolling mean, showing the underlying trend. The dashed blue line marks the statistically significant structural breakpoints, while the dashed red line denotes the 2024 U.S. election day: 05-Nov-2024. Table~\ref{tab:structural_breaks} presents the identified structural break dates with the corresponding test statistics for both blockchain and trading data of stablecoins and cryptocurrencies. These structural breaks suggest that the U.S. election introduced a shock to the market, leading to adjustments in both blockchain activity and trading patterns. The timing of these changes reflects how the market participants responded to political uncertainty.

% \vspace{0.5cm}  

% \textbf{Pre-election structural breaks in human-driven stablecoin transactions}

% \vspace{0.5cm}  

The transaction activity between EOA–EOA represents a direct measure of human-driven behavior on the blockchain. Unlike SC-SC transaction activity, which is run by the bots or automated algorithms, EOA–EOA interactions capture the real-time behavior of individual users, traders, and investors. As such, EOA–EOA transaction activity can be viewed as a behavioral signal, offering insights into how participants adjust their actions in response to election-induced events. In this section, we focus on the EOA–EOA blockchain transactions of USDT and USDC to examine whether election-induced events, i.e., the 2024 U.S. Presidential election, trigger structural changes in human-driven activity in the cryptocurrency ecosystem.

From Figs.~\ref{fig:USDT_HS_BTC}(a) and \ref{fig:USDC_HS_ETH}(a), we see that the BP test identified a significant structural breakpoint for EOA–EOA transactions on 03 November 2024, just two days before the U.S. Presidential election, for both USDT and USDC in human-driven EOA-EOA transactions. To validate the statistical significance of the identified breakpoints, we employed two  Bai-Perron SupF tests and AAFT surrogate analysis. First, we applied the Bai-Perron SupF test for detecting unknown break dates, where the null hypothesis assumes no structural break. The results reported in Table~\ref{tab:structural_breaks} strongly reject the null hypothesis for both stablecoins, with SupF statistics of 34.63 for USDC and 76.82 for USDT, both exceeding conventional significance thresholds. To further validate the robustness of these findings against autocorrelated noise, we conducted AAFT surrogate analysis. The results show p-values < 0.001 for both tokens, indicating that the probability of observing similar structural breaks within a ±3-day window by random chance is extremely low. This dual validation confirms that the identified breakpoints are both statistically significant and non-trivial. These results provide clear and robust evidence that both ERC-20 blockchain transaction USDT and USDC data experienced significant structural breaks around early November 2024, coinciding with the period just before the 2024 U.S. Presidential election. The consistent evidence confirms a robust structural break for both stablecoins, ruling out random fluctuations and indicating a significant shift in market behavior.

The timing of the identified break date using the BP test is critical, as it indicates a shift in user behavior in anticipation of the election outcome, rather than as a reaction to the announced results. The identified structural break in human-driven, EOA-to-EOA transactions suggests a fundamental change in the dynamics of stablecoin activity. This shift likely reflects a period of precautionary reallocation, wherein investors moved capital from more volatile cryptocurrencies into stablecoins as a hedge against potential election-induced uncertainty. This pattern aligns with established safe-haven behavior during periods of political stress \cite{wei2018impact,bianchi2020stablecoins}, where market participants enter a "wait-and-watch" phase and reduce exposure to high-risk assets. The pronounced response in human-driven activity further suggests that election-induced uncertainty exerted direct behavioral influence on individual participants, who typically demonstrate greater sensitivity to political and macroeconomic shocks.

\begin{table}[h]
\centering
\caption{Structural break test results for stablecoins and major cryptocurrencies. All breaks identified by the Bai–Perron SupF test show significant p-values, providing strong evidence against the null hypothesis of no structural break. AAFT surrogate testing (1000 iterations) confirms these breaks are non-trivial. $^{***}$ denotes statistical significance at the 0.001 level.}
\label{tab:structural_breaks}
\renewcommand{\arraystretch}{1.5}

\begin{tabular}{@{}p{2cm}p{2cm}p{2cm}p{2.5cm}p{4cm}@{}}
\toprule
\textbf{Asset} & \textbf{Break Date} & \textbf{SupF Statistic} & \textbf{AAFT p-value}  & \textbf{Comment}  \\
\midrule
\multicolumn{5}{c}{\textbf{Blockchain Data: EOA--EOA}} \\
\midrule
USDC & 2024-11-03 & 34.63$^{***}$ & <0.001 & Early human-driven response before election day.\\
USDT & 2024-11-03 & 76.82$^{***}$ & <0.001 & Early human-driven response before election day.\\
\midrule
\multicolumn{5}{c}{\textbf{Trading Data (Exchange)}} \\
\midrule
USDC-USD & 2024-11-05 & 79.04$^{***}$ & <0.001 & Immediate market reaction on election day.  \\
USDT-USD & 2024-11-05 & 315.23$^{***}$ & <0.001  & Immediate market reaction on election day. \\
BTC-USD  & 2024-11-09 & 3049.02$^{***}$ & <0.001 & Spillover to broader crypto markets after election. \\
ETH-USD  & 2024-11-06 & 137.60$^{***}$ & <0.001  & Spillover to broader crypto markets after election. \\
\midrule
\multicolumn{5}{c}{\textbf{Blockchain Data: SC--SC}} \\
\midrule
USDC & 2025-01-02 & 266.45$^{***}$ & <0.001 & Delayed adjustment in automated contract activity.\\
USDT & 2025-01-16 & 57.00$^{***}$ & <0.001 & Delayed adjustment in automated contract activity. \\
\botrule
\end{tabular}

\end{table}

\subsection{Hilbert Spectrum: extreme event in cryptocurrency}
\label{Result:HS}
The election-induced anticipatory shift in Human-driven EOA-EOA stablecoin blockchain transactions raises the question of whether it was subsequently followed by notable volatility in the broader cryptocurrency market. To investigate this, we employ the HHT, based on EMD, to assess whether the structural break observed in EOA–EOA transactions signals impending shifts in broader cryptocurrency market dynamics. This method provides a time–frequency representation of price behavior, enabling the detection of sudden changes in the cryptocurrency market during the election period. The HS, $H(t,\omega)$ is estimated by averaging across all IMFs derived from the closing price time series of BTC and ETH, as described in Section~\ref{Hilbert spectrum}. Equal weighting of IMFs ensures that the final spectrum captures the overall energy distribution without bias toward any single component.

Figures~\ref{fig:USDT_HS_BTC}(b) and \ref{fig:USDC_HS_ETH}(b) display the HS for BTC and ETH closing prices. Both reveal concentrated regions of high instantaneous energy, coinciding with the U.S. Presidential election. These intense energy patches signify abrupt market adjustments and increased volatility, corresponding to a positive extreme event as BTC and ETH prices rose sharply during this period. The normalized instantaneous energy $IE_N(t)$ plots in Figures~\ref{fig:USDT_HS_BTC}(c) and \ref{fig:USDC_HS_ETH}(c)  further support this interpretation. The peak in $IE_N(t)$ exceeds the statistical threshold $E_{th} = E_{\mu} + 4\sigma$, confirming that the election week triggered statistically significant extreme events. Prior studies identify $4\sigma$ exceedance as a hallmark of extreme events, and its occurrence here directly links the volatility burst to the election shock.
Along with the HS, the BP test detects structural breaks on 06-Nov-2024 for ETH and 09-Nov-2024 for BTC as present in Table~\ref{tab:structural_breaks}. These breakpoints align with the concentrated energy regions in the Hilbert spectra, reinforcing evidence of major structural adjustments in response to election-induced uncertainty. Overall, the sentiment shift captured by stablecoin blockchain activity on 03 November preceded the election-driven volatility, which subsequently manifested as structural breaks and positive extreme events in BTC and ETH, providing strong evidence of the predictive value of stablecoin flows for market-wide shocks.

\subsubsection{Structural breaks in stablecoin trading volumes}

\begin{figure}[H]
    \centering
	\includegraphics[width= 0.9\linewidth]{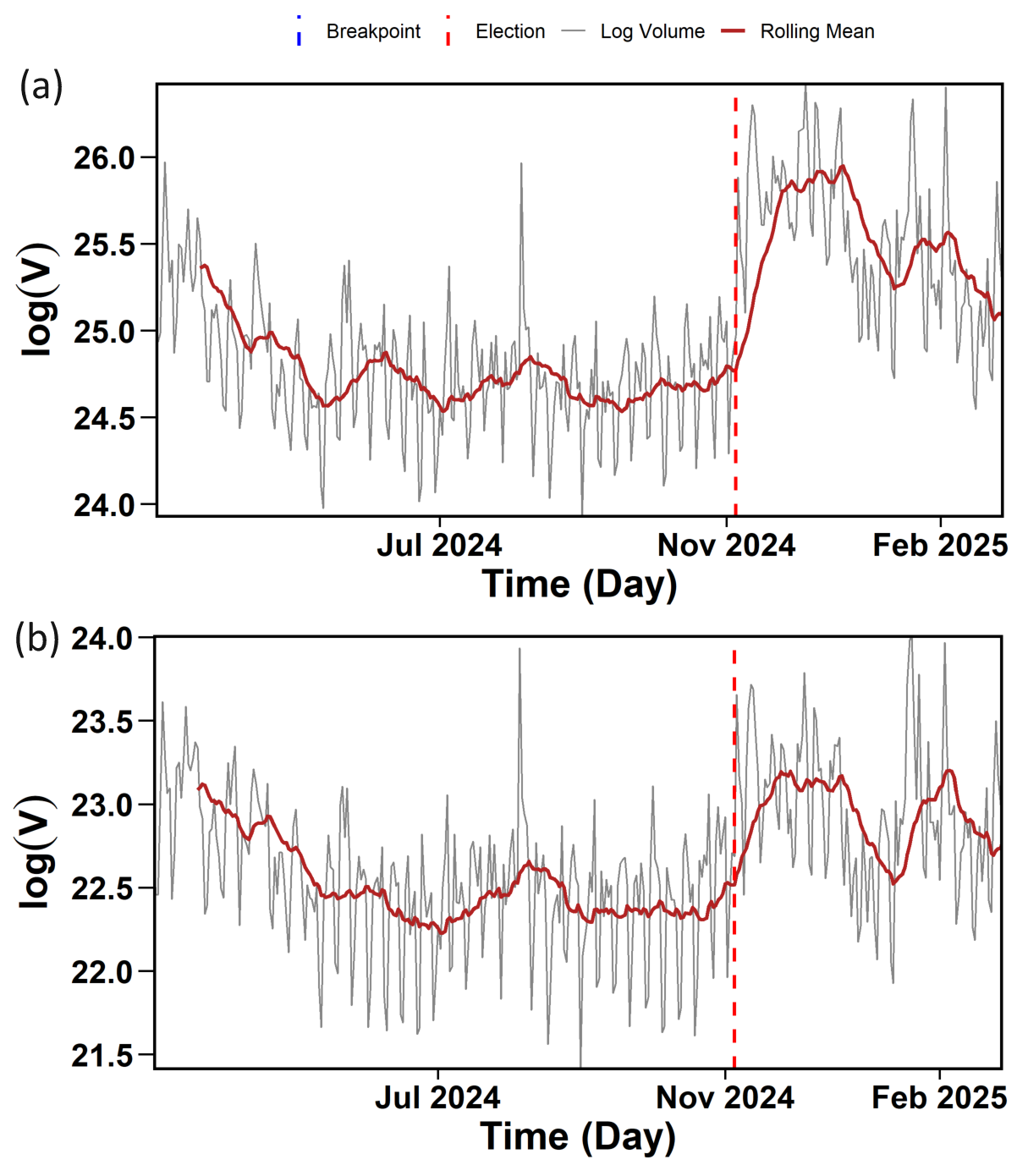}
    \caption{Structural break analysis of stablecoin trading volumes. The BP test detects significant structural breaks in (a) USDT and (b) USDC trading volumes on 5 November 2024, coinciding with the U.S. Presidential election. The red dashed line marks the election date, and the blue line would indicate the break date; however, because the two coincide, only the red dashed line is visible. These synchronized breaks confirm that the blockchain anticipatory signal was followed by a broad market adjustment in centralized exchange activity.}    \label{fig:SBA_Trading}
\end{figure}

% \begin{subfigure}[t]{0.48\textwidth}
%	\centering
%	\includegraphics[width=\linewidth]{USDT_trading.png}
%	\caption{USDT: Trading.}
%	\label{fig:usdt_break2}
%\end{subfigure}
%\hfill
%\begin{subfigure}[t]{0.48\textwidth}
%	\centering
%	\includegraphics[width=\linewidth]{USDC_trading.png}
%	\caption{USDC: Trading.}
%	\label{fig:usdc_break2}
%\end{subfigure}

The anticipatory shift in human-driven, EOA–EOA stablecoin transactions identified on 03-Nov-2024 served as a leading indicator. We now examine whether this blockchain signal was followed by a corresponding structural break in the broader market, as reflected in the trading volume on centralized exchanges (CEX). Figures~\ref{fig:SBA_Trading}(a) and~\ref{fig:SBA_Trading}(b) identified a structural break in the trading volume for both the USDT and USDC on 05-Nov-2024, the same day of the U.S. Presidential election. As shown in Table~\ref{tab:structural_breaks}, these breakpoints are statistically significant, with SupF statistics of 79.04 for USDC and 315.23 for USDT. The robustness of these findings is further confirmed by AAFT surrogate analysis, which gives p-values of $< 0.001$ for both stablecoins, indicating that the identified breaks are statistically significant. In this analysis, we use trading volume rather than USD price because stablecoins are pegged to the dollar; thus, volume is a more meaningful indicator of shifts in market activity and capital flows.

The sequence of structural breaks reveals an important pattern. The break in EOA–EOA transactions on 03-Nov-2024 indicates a precautionary shift, with investors moving capital to blockchain, likely to hedge against election uncertainty. This blockchain activity was the first measurable sign of changing user behavior. The subsequent break in CEX trading volume on 05-Nov-2024 marks the point of market-wide realization, where the unfolding election results triggered a surge in trading activity. Critically, this volatility subsequently spilled over into major cryptocurrencies. The BP test detects significant structural breaks on 06-Nov-2024 for ETH and 09-Nov-2024 for BTC, aligning with the period of intense market adjustment identified by the HS. This sequential pattern from a blockchain behavioral shift, to a surge in stablecoin trading and finally to structural breaks in major cryptocurrencies, demonstrates how election-induced anticipation culminated in a broad market reaction. The consistent sequence across stablecoins and major cryptocurrencies confirms the 2024 U.S. election was a significant structural shock and underscores the value of blockchain data as a leading indicator of sentiment. Together, these results further support the role of human-driven blockchain activity as an effective early warning system, revealing emerging market stress days before it becomes visible in exchange trading or price dynamics.

\subsection{Structural breaks in automated SC–SC transactions}
\begin{figure}[H]
\centering
\includegraphics[width= 0.9\linewidth]{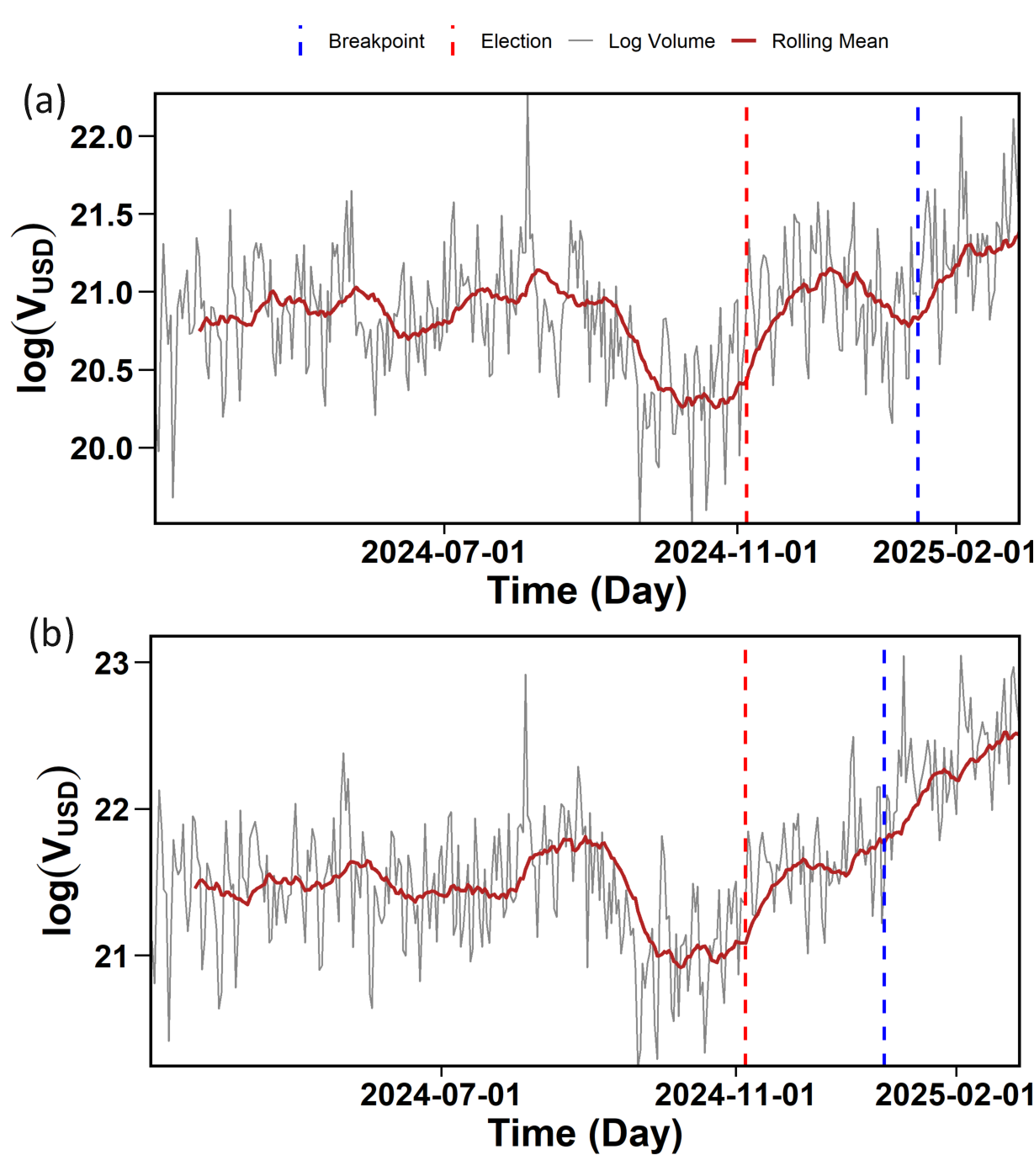}	\caption{Structural Break Analysis of Automated SC–SC Stablecoin Transactions. The Bai–Perron test identifies significant post-election structural breaks in automated SC–SC transaction volumes for (a) USDT on 16 January 2025,  and (b) USDC on 2 January 2025. The red dashed line marks the 2024 U.S. election date, while the blue line indicates the respective break dates. These delayed adjustments indicate that bot-driven systems responded only after market stabilization, contrasting with the immediate, human-driven shifts observed before the election.}	\label{fig:SBA_SC-SC}
\end{figure}

In this section, we discuss how the automated, code-driven SC-SC blockchain transaction responded to the election-induced effect different way than a human-driven one. It is essential to distinguish the nature of these transactions. SC–SC interactions represent automated, pre-programmed operations executed by bots without direct human intervention, whereas EOA–EOA transactions are initiated by human users from their personal wallets. As shown in Figures~\ref{fig:SBA_SC-SC}(a) and \ref{fig:SBA_SC-SC}(b), the BP test identified that the structural breaks for USDT and USDC occurred months after the election, on 16-Jan-2025 and 02-Jan-2025, respectively. These breaks are statistically robust, with high SupF statistics of 57.00 for USDT and 266.45 for USDC. The identified break date is also confirmed by AAFT surrogate testing with $< 0.001$, as detailed in Table~\ref{tab:structural_breaks}. This delayed response shows a clear difference from the immediate, precautionary shift seen in human-driven (EOA–EOA) transactions just before the election. This timing difference indicates a key finding that the 2024 U.S. election was primarily a behavioral shock that immediately affected human participants, but only later triggered adjustments in automated systems. The January 2025 breaks likely reflect algorithmic recalibrations to the new, post-election market environment, such as updated trading strategies or protocol changes, once the new market conditions have stabilized. This confirms that bot-driven SC–SC activity is less sensitive to sudden exogenous political events in real-time, instead reacting to more established trends. These results clearly demonstrate the fundamental difference between human-driven and code-driven activity, as humans react in real time and exhibit herding behaviour that cannot be replicated by automated SC. This result strengthens the importance of our earlier findings. It shows that human-driven EOA–EOA blockchain activity is a unique and valuable indicator, capturing immediate sentiment shifts that precede both exchange trading volumes and the subsequent adjustments of automated systems.

\subsection{Structural VAR-based analysis}
\label{Result:SVAR}

In this section, we study whether the 2024 U.S. Presidential election triggered a regime shift in stablecoin markets by analyzing volatility spillovers between USDT and USDC. To examine this, we employ SVAR analysis to investigate how election-induced uncertainty affected the dynamic relationships between stablecoins. This framework enables us to move beyond simple correlations to assess both the magnitude and direction of causal volatility spillovers between USDT and USDC. To identify the structural model, we utilize a Cholesky decomposition~\cite{pourahmadi2007cholesky}, which requires imposing a specific ordering on the variables to orthogonalize the shocks. We present results under two distinct identification schemes to test the robustness of the spillover dynamics and to understand the sensitivity of our conclusions to different assumptions about market leadership. The analysis is conducted using both blockchain transaction data, EOA-EOA transactions, and exchange-based trading volume data. This ensures that it captures spillover dynamics across different market segments. To assess changes over time, we examine the SVAR across three periods: the full sample, the pre-election period defined as up to 04-Nov-2024, and the post-election period defined as from 05-Nov-2024 onward. This division allows for a direct comparison of relationship dynamics before and after the election event.

\begin{table}[h]
\centering
\caption{Wald test results for structural breaks in blockchain and trading data. The high test statistics (p < 0.0001) provide strong evidence of regime change, with trading data displaying substantially larger breaks (Wald = 5381.56) than the blockchain transaction series.}
\label{tab:wald_results_combined}
\renewcommand{\arraystretch}{1.5}

\begin{tabular}{@{}p{3cm}p{2cm}p{2cm}p{3cm}@{}}
\toprule
\textbf{Identification Ordering} & \textbf{Wald Statistic} & \textbf{P-value} & \textbf{Decision}  \\
\midrule
\multicolumn{4}{c}{\textbf{Blockchain Data (EOA--EOA)}} \\
\midrule
USDC-First & 295.21  & $<$0.0001  & Significant regime change.   \\
USDT-First & 138.78  & $<$0.0001   & Significant regime change.  \\
\midrule
\multicolumn{4}{c}{\textbf{Trading Data}} \\
\midrule
USDT-First & 876.24  & $<$0.0001  & Significant regime change.  \\
USDC-First & 5381.56 & $<$0.0001  & Significant regime change.  \\
\botrule
\end{tabular}

\end{table}

\begin{table}[h]
\centering
\caption{Impact matrices for blockchain and trading data under alternative identification schemes. All values are reported in their original 0-1 scale and are not multiplied by 100. The post-election period shows a 28–48\% increase in volatility spillovers, with USDT emerging as the dominant transmission channel of election-induced stress across both blockchain and exchange markets.}
\label{tab:impact_matrices_combined}
\renewcommand{\arraystretch}{1.5}
\
\begin{tabular}{@{}p{3cm}p{2cm}p{2cm}p{2cm}@{}}
\toprule
\textbf{Response} & \textbf{Full Sample} & \textbf{Pre-Election} & \textbf{Post-Election} \\
\midrule
\multicolumn{4}{c}{\textbf{Blockchain Data (EOA--EOA): USDC Precedes USDT}} \\
\midrule
USDC to USDC Shock & 0.3441 & 0.3232 & 0.4410 \\
USDT to USDC Shock & 0.1881 & 0.1620 & 0.2399 \\
USDT to USDT Shock & 0.1952 & 0.1645 & 0.2334 \\
\midrule
\multicolumn{4}{c}{\textbf{Blockchain Data (EOA--EOA): USDT Precedes USDC}} \\
\midrule
USDT to USDT Shock & 0.2711 & 0.2309 & 0.3347 \\
USDC to USDT Shock & 0.2388 & 0.2268 & 0.3161 \\
USDC to USDC Shock & 0.2478 & 0.2303 & 0.3075 \\
\midrule
\multicolumn{4}{c}{\textbf{Trading Data: USDT Precedes USDC}} \\
\midrule
USDT to USDT Shock & 0.2474 & 0.2175 & 0.2782 \\
USDC to USDT Shock & 0.2632 & 0.2373 & 0.3077 \\
USDC to USDC Shock & 0.1042 & 0.0969 & 0.1133 \\
\midrule
\multicolumn{4}{c}{\textbf{Trading Data: USDC Precedes USDT}} \\
\midrule
USDC to USDC Shock & 0.2831 & 0.2564 & 0.3279 \\
USDT to USDC Shock & 0.2300 & 0.2013 & 0.2611 \\
USDT to USDT Shock & 0.0911 & 0.0822 & 0.0962 \\
\botrule
\end{tabular}

\end{table}

Table~\ref{tab:impact_matrices_combined} presents the estimated impact matrices. The ordering assumption implies that a variable placed first can contemporaneously affect the second, but the second can only affect the first with a lag. The results show that all elements increase in the post-election period for both blockchain and trading data and under both identification orderings. This indicates higher volatility and stronger spillover effects. A closer look at the values reveals that USDT’s responses are particularly notable. For instance, in blockchain data under the USDT-first ordering, the USDT own-shock effect jumps from 0.2309 to 0.3347, while in trading data, it rises from 0.2175 to 0.2782. Cross-effects are also stronger when USDT is the leading variable, for example, USDC to USDT shocks in trading grow from 0.2373 to 0.3077. The own-shock effects rise by 28–45\% in the post-election period compared with the pre-election period, suggesting a significant increase in intrinsic volatility following the election. The cross-stablecoin spillover effects also rise sharply in the post-election period, with the contemporaneous spillover parameters increasing by 39–48\% across both data sources and identification orderings. These results suggest that USDT plays a disproportionately important role in driving post-election volatility. This pattern aligns with the fact that USDT exhibits higher market capitalization and deeper liquidity as compared to USDC. The findings, therefore, indicate that election-induced stress was transmitted more strongly through USDT, amplifying its role as the primary channel of spillovers in stablecoin markets. Table~\ref{tab:wald_results_combined} reports the Wald test results for both blockchain EOA–EOA and trading data. The results show large and highly significant statistics with $p$ < 0.0001, confirming clear structural breaks in blockchain data and even stronger shifts in trading data, suggesting that election-induced changes were more pronounced in exchange-based activity. The SVAR results confirm a clear regime change in stablecoin dynamics following the 2024 U.S. Presidential election, characterized by amplified volatility and deeper interconnectedness between USDT and USDC after the election. Together with the structural break tests, these findings provide strong evidence that the 2024 U.S. Presidential election triggered fundamental regime shifts in the cryptocurrency ecosystem. They indicate that political uncertainty can temporarily alter the transmission of volatility, tighten linkages between stablecoins, and accelerate the spread of shocks across markets. These patterns also suggest that shifts in blockchain activity and trading responses form a coordinated adjustment process during major political events.

\section{Conclusion and Discussion}
\label{conclu}

This study establishes that human-driven blockchain transaction activity serves as an early warning system for financial market stress during periods of political uncertainty. The 2024 U.S. Presidential election provided a clear natural experiment that revealed a sequential pattern of reactions across the cryptocurrency ecosystem. Our analysis demonstrates that human-driven activity, reflected in EOA–EOA stablecoin blockchain transfers, was the first to react to political uncertainty, with structural breaks identified on November 3, 2024, two days before the election. This signals a precautionary shift into stablecoins ahead of anticipated volatility due to the election. These early human-led adjustments, confirmed through Amplitude-Adjusted Fourier Transform surrogate testing with p < 0.001, emerged before any breaks appeared in exchange-based trading activity or automated SC–SC transactions. This sequence shows that human-driven blockchain behavior provides a clear early warning signal, preceding reactions observed in centralized markets and algorithmic systems. The subsequent structural breaks in exchange-based trading volumes on election day (November 5) and the extreme price movements in BTC and ETH identified through Hilbert Spectrum analysis, with both exceeding the $E_{th} = E_{\mu} + 4\sigma$ extreme event threshold, confirm that these initial blockchain early signals accurately anticipated broader market turbulence. A clear behavioral divergence was observed between human and algorithmic responses to the political stress. Human-driven activity adjusted rapidly to pre-election uncertainty, whereas automated SC–SC transactions exhibited delayed responses, with structural breaks appearing only in January 2025. This pattern suggests that human sentiment and intentional decisions dominate short-term adjustments, while algorithmic systems adapt more gradually once new equilibrium conditions emerge. Structural vector autoregression analysis confirms a post-election regime shift, with volatility and spillover effects between USDT and USDC rising by 28–48\% relative to the pre-election period. Wald test results with p < 0.0001 verify the statistical significance of this shift, underscoring how political events can reshape stablecoin dynamics.  

Overall, these findings demonstrate that distinguishing between human and algorithmic blockchain behavior provides an early warning framework for market stress. Human-driven EOA–EOA activity reacts first to political uncertainty and offers advanced signals that appear before changes in exchange trading or automated smart contract activity. This early response improves our understanding of how political shocks transmit through decentralized markets and reveals a clear sequence in behavioral adjustments. These findings also have important practical implications for investors, risk managers, and policymakers who must make decisions during periods of political uncertainty in cryptocurrency markets. For investors, early human-driven signals provide clear, data-backed guidance that can help them adjust their portfolios before volatility fully unfolds. For risk managers, integrating real-time blockchain activity strengthens short-term stress monitoring and improves the accuracy of market risk models, and policymakers can gain clearer insights into how political events influence stablecoin flows and financial stability.

Our analysis is limited to ERC-20 blockchain transactions and does not capture stablecoin activity occurring on other blockchain platforms. In the future, we plan to extend this analysis across multiple blockchain ecosystems, incorporate machine learning techniques for regime detection, and examine whether similar early warning patterns emerge during other geopolitical or macroeconomic events.

\bmhead{Acknowledgements}
We acknowledge NIT Sikkim for providing doctoral fellowships to Kundan Mukhia, Salam Rbindrajit Luwang and Buddha Nath Sharma.

% \section*{Declarations}

% Some journals require declarations to be submitted in a standardised format. Please check the Instructions for Authors of the journal to which you are submitting to see if you need to complete this section. If yes, your manuscript must contain the following sections under the heading `Declarations':

% \begin{itemize}
% \item Conflict of interest/Competing interests (check journal-specific guidelines for which heading to use)
% \item Ethics approval and consent to participate
% \item Consent for publication
% \item Materials availability
% \end{itemize}

% \noindent
% If any of the sections are not relevant to your manuscript, please include the heading and write `Not applicable' for that section. 

\bibliography{sn-bibliography}

@article{mahata2020identification,
  title={Identification of short-term and long-term time scales in stock markets and effect of structural break},
  author={Mahata, Ajit and Bal, Debi Prasad and Nurujjaman, Md},
  journal={Physica A: Statistical Mechanics and its Applications},
  volume={545},
  pages={123612},
  year={2020},
  publisher={Elsevier}
}

@article{sengupta2025forecasting,
  title={Forecasting CPI inflation under economic policy and geopolitical uncertainties},
  author={Sengupta, Shovon and Chakraborty, Tanujit and Singh, Sunny Kumar},
  journal={International Journal of Forecasting},
  volume={41},
  number={3},
  pages={953--981},
  year={2025},
  publisher={Elsevier}
}

@article{besher5705079modeling,
  title={Modeling US Climate Policy Uncertainty: From Causal Identification to Probabilistic Forecasting},
  author={Besher, Donia and Sengupta, Anirban and Chakraborty, Tanujit},
  journal={Available at SSRN 5705079},
  year={2025}
}

@article{chakraborty2025neural,
  title={Neural ARFIMA model for forecasting BRIC exchange rates with long memory under oil shocks and policy uncertainties},
  author={Chakraborty, Tanujit and Besher, Donia and Panja, Madhurima and Sengupta, Shovon},
  journal={arXiv preprint arXiv:2509.06697},
  year={2025}
}

@article{chaudhuri2003random,
  title={Random walk versus breaking trend in stock prices: Evidence from emerging markets},
  author={Chaudhuri, Kausik and Wu, Yangru},
  journal={Journal of Banking \& Finance},
  volume={27},
  number={4},
  pages={575--592},
  year={2003},
  publisher={Elsevier}
}

@article{worthington2007gold,
  title={Gold investment as an inflationary hedge: Cointegration evidence with allowance for endogenous structural breaks},
  author={Worthington, Andrew C and Pahlavani, Mosayeb},
  journal={Applied Financial Economics Letters},
  volume={3},
  number={4},
  pages={259--262},
  year={2007},
  publisher={Taylor \& Francis}
}

@article{bai2003computation,
  title={Computation and analysis of multiple structural change models},
  author={Bai, Jushan and Perron, Pierre},
  journal={Journal of applied econometrics},
  volume={18},
  number={1},
  pages={1--22},
  year={2003},
  publisher={Wiley Online Library}
}

@article{bai1998estimating,
  title={Estimating and testing linear models with multiple structural changes},
  author={Bai, Jushan and Perron, Pierre},
  journal={Econometrica},
  pages={47--78},
  year={1998},
  publisher={JSTOR}
}

@article{amarasinghe2015dynamic,
  title={Dynamic relationship between interest rate and stock price: Empirical evidence from colombo stock exchange},
  author={Amarasinghe, AA},
  journal={International Journal of Business and Social Science},
  volume={6},
  number={4},
  year={2015},
  publisher={Centre for Promoting Ideas, USA}
}

@article{said1984testing,
  title={Testing for unit roots in autoregressive-moving average models of unknown order},
  author={Said, Said E and Dickey, David A},
  journal={Biometrika},
  volume={71},
  number={3},
  pages={599--607},
  year={1984},
  publisher={Oxford University Press}
}

@article{herranz2017unit,
  title={Unit root tests},
  author={Herranz, Edward},
  journal={Wiley Interdisciplinary Reviews: Computational Statistics},
  volume={9},
  number={3},
  pages={e1396},
  year={2017},
  publisher={Wiley Online Library}
}

@article{paparoditis2018asymptotic,
  title={The asymptotic size and power of the augmented Dickey--Fuller test for a unit root},
  author={Paparoditis, Efstathios and Politis, Dimitris N},
  journal={Econometric Reviews},
  volume={37},
  number={9},
  pages={955--973},
  year={2018},
  publisher={Taylor \& Francis}
}

@article{kim2017unit,
  title={Unit roots in economic and financial time series: A re-evaluation at the decision-based significance levels},
  author={Kim, Jae H and Choi, In},
  journal={Econometrics},
  volume={5},
  number={3},
  pages={41},
  year={2017},
  publisher={MDPI}
}

@article{zhang2009estimating,
  title={Estimating the impact of extreme events on crude oil price: An EMD-based event analysis method},
  author={Zhang, Xun and Yu, Lean and Wang, Shouyang and Lai, Kin Keung},
  journal={Energy Economics},
  volume={31},
  number={5},
  pages={768--778},
  year={2009},
  publisher={Elsevier}
}

@misc{cutler1988moves,
  title={What moves stock prices?},
  author={Cutler, David M and Poterba, James M and Summers, Lawrence H},
  year={1988},
  publisher={National Bureau of Economic Research Cambridge, Mass., USA}
}

@article{beber2010cannot,
  title={When it cannot get better or worse: The asymmetric impact of good and bad news on bond returns in expansions and recessions},
  author={Beber, Alessandro and Brandt, Michael W},
  journal={Review of Finance},
  volume={14},
  number={1},
  pages={119--155},
  year={2010},
  publisher={Oxford University Press}
}

@article{corbet2020impact,
  title={The impact of macroeconomic news on Bitcoin returns},
  author={Corbet, Shaen and Larkin, Charles and Lucey, Brian M and Meegan, Andrew and Yarovaya, Larisa},
  journal={The European Journal of Finance},
  volume={26},
  number={14},
  pages={1396--1416},
  year={2020},
  publisher={Taylor \& Francis}
}

@article{bloom2009impact,
  title={The impact of uncertainty shocks},
  author={Bloom, Nicholas},
  journal={econometrica},
  volume={77},
  number={3},
  pages={623--685},
  year={2009},
  publisher={Wiley Online Library}
}

@article{bianchi2020stablecoins,
  title={Stablecoins and cryptocurrency returns: What is the role of Tether?},
  author={Bianchi, Daniele and Rossini, Luca and Iacopini, Matteo},
  journal={Available at SSRN 3605451},
  year={2020}
}

@article{wei2018impact,
  title={The impact of Tether grants on Bitcoin},
  author={Wei, Wang Chun},
  journal={Economics Letters},
  volume={171},
  pages={19--22},
  year={2018},
  publisher={Elsevier}
}

@article{paramati2011empirical,
  title={An empirical analysis of stock market performance and economic growth: Evidence from India},
  author={Paramati, Sudharshan Reddy and Gupta, Rakesh},
  journal={International Research Journal of Finance and Economics},
  year={2011},
  publisher={EuroJournals}
}

@article{yang2020novel,
  title={A novel two-stage approach for cryptocurrency analysis},
  author={Yang, Boyu and Sun, Yuying and Wang, Shouyang},
  journal={International Review of Financial Analysis},
  volume={72},
  pages={101567},
  year={2020},
  publisher={Elsevier}
}

@article{dickey1981likelihood,
  title={Likelihood ratio statistics for autoregressive time series with a unit root},
  author={Dickey, David A and Fuller, Wayne A},
  journal={Econometrica: journal of the Econometric Society},
  pages={1057--1072},
  year={1981},
  publisher={JSTOR}
}

@article{bai2003critical,
  title={Critical values for multiple structural change tests},
  author={Bai, Jushan and Perron, Pierre},
  journal={The Econometrics Journal},
  volume={6},
  number={1},
  pages={72--78},
  year={2003},
  publisher={Oxford University Press Oxford, UK}
}

@article{huang1998empirical,
  title={The empirical mode decomposition and the Hilbert spectrum for nonlinear and non-stationary time series analysis},
  author={Huang, Norden E and Shen, Zheng and Long, Steven R and Wu, Manli C and Shih, Hsing H and Zheng, Quanan and Yen, Nai-Chyuan and Tung, Chi Chao and Liu, Henry H},
  journal={Proceedings of the Royal Society of London. Series A: mathematical, physical and engineering sciences},
  volume={454},
  number={1971},
  pages={903--995},
  year={1998},
  publisher={The Royal Society}
}

@article{huang2003applications,
  title={Applications of Hilbert--Huang transform to non-stationary financial time series analysis},
  author={Huang, Norden E and Wu, Man-Li and Qu, Wendong and Long, Steven R and Shen, Samuel SP},
  journal={Applied stochastic models in business and industry},
  volume={19},
  number={3},
  pages={245--268},
  year={2003},
  publisher={Wiley Online Library}
}

@book{kilian2017structural,
  title={Structural vector autoregressive analysis},
  author={Kilian, Lutz and L{\"u}tkepohl, Helmut},
  year={2017},
  publisher={Cambridge University Press},
  address={Cambridge} 
}

@misc{etherscan2025eth,
  title        = {Etherscan: Ethereum Blockchain Explorer and Analytics Platform},
  author       = {{Etherscan}},
  howpublished = {\url{https://etherscan.io/}},
  
  year         = {2025}
}

@misc{coinmarketcap2025stablecoin,
  title        = {Top Stablecoin Tokens by Market Capitalization},
  author       = {{CoinMarketCap}},
  howpublished = {\url{https://coinmarketcap.com/view/stablecoin/}},
  
  year         = {2025}
}

@inproceedings{chen2020traveling,
  title={Traveling the token world: A graph analysis of ethereum erc20 token ecosystem},
  author={Chen, Weili and Zhang, Tuo and Chen, Zhiguang and Zheng, Zibin and Lu, Yutong},
  booktitle={Proceedings of The Web Conference 2020},
  pages={1411--1421},
  year={2020}
}

@article{szabo1997idea,
  title={The idea of smart contracts},
  author={Szabo, Nick},
  journal={Nick Szabo’s papers and concise tutorials},
  volume={6},
  number={1},
  pages={199},
  year={1997}
}

@incollection{kolvart2016smart,
  title={Smart contracts},
  author={Kolvart, Merit and Poola, Margus and Rull, Addi},
  booktitle={The Future of Law and etechnologies},
  pages={133--147},
  year={2016},
  publisher={Springer},
  address={Cham}
}

@misc{wikipedia2025unixtime,
  title        = {Unix time},
  author       = {{Wikipedia contributors}},
  howpublished = {\url{https://en.wikipedia.org/wiki/Unix_time}},
  year         = {2025}
}

@article{omrane2025exploring,
  title={Exploring Volatility Reactions in Cryptocurrency Markets Using Intraday Macroeconomic News Analysis},
  author={Omrane, Walid Ben and Dabbou, Halim and Saadi, Samir and Savaser, Tanseli and Sebai, Saber},
  journal={International Review of Economics \& Finance},
  pages={104509},
  year={2025},
  publisher={Elsevier}
}

@article{corbet2020cryptocurrency,
  title={Cryptocurrency reaction to fomc announcements: Evidence of heterogeneity based on blockchain stack position},
  author={Corbet, Shaen and Larkin, Charles and Lucey, Brian and Meegan, Andrew and Yarovaya, Larisa},
  journal={Journal of Financial Stability},
  volume={46},
  pages={100706},
  year={2020},
  publisher={Elsevier}
}

@article{chen2020blockchain,
  title={Blockchain disruption and decentralized finance: The rise of decentralized business models},
  author={Chen, Yan and Bellavitis, Cristiano},
  journal={Journal of Business Venturing Insights},
  volume={13},
  pages={e00151},
  year={2020},
  publisher={Elsevier}
}

@article{dyhrberg2016bitcoin,
  title={Bitcoin, gold and the dollar--A GARCH volatility analysis},
  author={Dyhrberg, Anne Haubo},
  journal={Finance research letters},
  volume={16},
  pages={85--92},
  year={2016},
  publisher={Elsevier}
}

@article{katsiampa2017volatility,
  title={Volatility estimation for Bitcoin: A comparison of GARCH models},
  author={Katsiampa, Paraskevi},
  journal={Economics letters},
  volume={158},
  pages={3--6},
  year={2017},
  publisher={Elsevier}
}

@article{ante2021influence,
  title={The influence of stablecoin issuances on cryptocurrency markets},
  author={Ante, Lennart and Fiedler, Ingo and Strehle, Elias},
  journal={Finance Research Letters},
  volume={41},
  pages={101867},
  year={2021},
  publisher={Elsevier}
}

@article{lyons2023keeps,
  title={What keeps stablecoins stable?},
  author={Lyons, Richard K and Viswanath-Natraj, Ganesh},
  journal={Journal of International Money and Finance},
  volume={131},
  pages={102777},
  year={2023},
  publisher={Elsevier}
}

@article{mackinnon1996numerical,
  title={Numerical distribution functions for unit root and cointegration tests},
  author={MacKinnon, James G},
  journal={Journal of applied econometrics},
  volume={11},
  number={6},
  pages={601--618},
  year={1996},
  publisher={Wiley Online Library}
}

@article{pastor2013political,
  title={Political uncertainty and risk premia},
  author={P{\'a}stor, L'ubo{\v{s}} and Veronesi, Pietro},
  journal={Journal of financial Economics},
  volume={110},
  number={3},
  pages={520--545},
  year={2013},
  publisher={Elsevier}
}

@article{shiller2020narrative,
  title={Narrative economics: How stories go viral and drive major economic events},
  author={Shiller, Robert J},
  year={2020},
  publisher={Princeton University Press}
}

@article{jabeur2025crypto,
  title={‘Crypto president’: Do narrative political signals drive cryptocurrency returns?},
  author={Jabeur, Sami Ben and Dhifaoui, Zouhaier and Bakkar, Yassine and Ballouk, Houssein},
  journal={Finance Research Letters},
  volume={78},
  pages={107194},
  year={2025},
  publisher={Elsevier}
}

@article{bialkowski2022high,
  title={High policy uncertainty and low implied market volatility: An academic puzzle?},
  author={Bia{\l}kowski, J{\k{e}}drzej and Dang, Huong Dieu and Wei, Xiaopeng},
  journal={Journal of Financial Economics},
  volume={143},
  number={3},
  pages={1185--1208},
  year={2022},
  publisher={Elsevier}
}

@article{lee2025stablecoin,
  title={Stablecoin depegging risk prediction},
  author={Lee, Yi-Hsi and Chiu, Yu-Fen and Hsieh, Ming-Hua},
  journal={Pacific-Basin Finance Journal},
  volume={90},
  pages={102640},
  year={2025},
  publisher={Elsevier}
}

@article{de2023intelligent,
  title={Intelligent design: stablecoins (in) stability and collateral during market turbulence},
  author={De Blasis, Riccardo and Galati, Luca and Webb, Alexander and Webb, Robert I},
  journal={Financial Innovation},
  volume={9},
  number={1},
  pages={85},
  year={2023},
  publisher={Springer}
}

@article{gregory2024break,
  title={Break a peg! A study of stablecoin co-instability},
  author={Gregory, Gadzinski and Alessio, Castello and Vito, Liuzzi and Patrice, Sargenti},
  journal={International Review of Financial Analysis},
  volume={96},
  pages={103608},
  year={2024},
  publisher={Elsevier}
}

@article{eichengreen2025stablecoin,
  title={Stablecoin devaluation risk},
  author={Eichengreen, Barry and Nguyen, My T and Viswanath-Natraj, Ganesh},
  journal={The European Journal of Finance},
  pages={1--28},
  year={2025},
  publisher={Taylor \& Francis}
}

@article{mukhia2024complex,
  title={Complex network analysis of cryptocurrency market during crashes},
  author={Mukhia, Kundan and Rai, Anish and Luwang, SR and Nurujjaman, Md and Majhi, Sushovan and Hens, Chittaranjan},
  journal={Physica A: Statistical Mechanics and its Applications},
  volume={653},
  pages={130095},
  year={2024},
  publisher={Elsevier}
}

@article{grobys2022tether,
  title={When tether says “Jump!” Bitcoin asks “How low?”},
  author={Grobys, Klaus and Huynh, Toan Luu Duc},
  journal={Finance Research Letters},
  volume={47},
  pages={102644},
  year={2022},
  publisher={Elsevier}
}

@article{sims1980macroeconomics,
  title={Macroeconomics and reality},
  author={Sims, Christopher A},
  journal={Econometrica: journal of the Econometric Society},
  pages={1--48},
  year={1980},
  publisher={JSTOR}
}

@book{lutkepohl2013introduction,
  title={Introduction to Multiple Time Series Analysis},
  author={L{\"u}tkepohl, Helmut},
  year={2013},
  publisher={Springer Science \& Business Media},
  address={Berlin}
}

@incollection{stock2016dynamic,
  title={Dynamic factor models, factor-augmented vector autoregressions, and structural vector autoregressions in macroeconomics},
  author={Stock, James H. and Watson, Mark W.},
  booktitle={Handbook of Macroeconomics},
  editor={Taylor, John B. and Uhlig, Harald},
  series={Handbook of Macroeconomics},
  volume={2},
  pages={415--525},
  year={2016},
  publisher={Elsevier},
  address={Elsevier}
 
}

@article{bernanke2005measuring,
  title={Measuring the effects of monetary policy: a factor-augmented vector autoregressive (FAVAR) approach},
  author={Bernanke, Ben S and Boivin, Jean and Eliasz, Piotr},
  journal={The Quarterly journal of economics},
  volume={120},
  number={1},
  pages={387--422},
  year={2005},
  publisher={MIT Press}
}

@article{schreiber2000surrogate,
  title={Surrogate time series},
  author={Schreiber, Thomas and Schmitz, Andreas},
  journal={Physica D: Nonlinear Phenomena},
  volume={142},
  number={3-4},
  pages={346--382},
  year={2000},
  publisher={Elsevier}
}

@article{lancaster2018surrogate,
  title={Surrogate data for hypothesis testing of physical systems},
  author={Lancaster, Gemma and Iatsenko, Dmytro and Pidde, Aleksandra and Ticcinelli, Valentina and Stefanovska, Aneta},
  journal={Physics Reports},
  volume={748},
  pages={1--60},
  year={2018},
  publisher={Elsevier}
}

@article{theiler1992testing,
  title={Testing for nonlinearity in time series: the method of surrogate data},
  author={Theiler, James and Eubank, Stephen and Longtin, Andr{\'e} and Galdrikian, Bryan and Farmer, J Doyne},
  journal={Physica D: Nonlinear Phenomena},
  volume={58},
  number={1-4},
  pages={77--94},
  year={1992},
  publisher={Elsevier}
}

@article{kugiumtzis1999test,
  title={Test your surrogate data before you test for nonlinearity},
  author={Kugiumtzis, D},
  journal={Physical Review E},
  volume={60},
  number={3},
  pages={2808},
  year={1999},
  publisher={APS}
}

@article{mukhia2025universal,
  title={Universal Patterns in the Blockchain: Analysis of EOAs and Smart Contracts in ERC20 Token Networks},
  author={Mukhia, Kundan and Luwang, SR and Nurujjaman, Md and Chakraborty, Tanujit and Saha, Suman and Hens, Chittaranjan},
  journal={arXiv preprint arXiv:2508.04671},
  year={2025}
}

@article{wald1943tests,
  title={Tests of statistical hypotheses concerning several parameters when the number of observations is large},
  author={Wald, Abraham},
  journal={Transactions of the American Mathematical society},
  volume={54},
  number={3},
  pages={426--482},
  year={1943},
  publisher={JSTOR}
}

@misc{xblock2025eth,
  title        = {XBlock: Ethereum Token and Transaction Data Explorer},
  author       = {{XBlock Team}},
  howpublished = {\url{https://xblock.pro/xblock-eth.html}},
  year         = {2025}
}

@article{pourahmadi2007cholesky,
  title={Cholesky decompositions and estimation of a covariance matrix: orthogonality of variance--correlation parameters},
  author={Pourahmadi, Mohsen},
  journal={Biometrika},
  volume={94},
  number={4},
  pages={1006--1013},
  year={2007},
  publisher={Oxford University Press}
}

\end{document}